\documentclass[useAMS,usenatbib]{mn2e}
\usepackage[dvips]{graphicx}
\usepackage{float}
\usepackage{amsmath}	
\usepackage{amssymb}	
\usepackage{multicol}   
\usepackage{bm}		    
\usepackage{pdflscape}	
\usepackage{array}
\usepackage{booktabs}
\usepackage{tabularx}
\usepackage{natbib}
\usepackage{rotating}
\usepackage{subfigure}
\title[Monthly Notices: \LaTeXe\ guide for authors]
  {Photometric Redshift Estimation of BASS DR3 Quasars by Machine Learning}
\author[C. Li et~al.]
  {Changhua Li$^{1,2,3}$,
  	Yanxia Zhang$^{1,4}$\thanks{Email: zyx@bao.ac.cn},
  	Chenzhou Cui$^{1,3}$\thanks{Email: ccz@bao.ac.cn}, Dongwei Fan$^{1,3}$, \and Yongheng Zhao$^{1}$, Xue-Bing Wu$^{5,6}$, Jing-Yi Zhang$^{1,4}$, Jun Han$^{1,3}$, \and Yunfei Xu$^{1,3}$, Yihan Tao$^{1,3}$, Shanshan Li$^{1,2,3}$, Boliang He$^{1,2,3}$ \\
  	$^1$ National Astronomical Observatories, Beijing, 100101, China\\
  	$^2$ University of Chinese Academy of Sciences, Beijing 100049, China\\
  	$^3$ National Astronomical Data Center, Beijing 100101, China\\
    $^4$ CAS Key Laboratory of Optical Astronomy, National Astronomical
Observatories, Beijing, 100101, China\\
  	$^5$ Department of Astronomy, School of Physics, Peking University, Beijing 100871, China\\
  	$^6$ Kavli Institute for Astronomy and Astrophysics, Peking University, Beijing 100871, China}
\date{Released 2002 Xxxxx XX}

\begin{document}

\label{firstpage}

\maketitle

\begin{abstract}
Correlating BASS DR3 catalogue with ALLWISE database, the data from optical and infrared information are obtained. The quasars from SDSS are taken as training and test samples while those from LAMOST are considered as external test sample. We propose two schemes to construct the redshift estimation models with XGBoost, CatBoost and Random forest. One scheme (namely one-step model) is to predict photometric redshifts directly based on the optimal models created by these three algorithms; the other scheme (namely two-step model) is to firstly classify the data into low- and high- redshift datasets, and then predict photometric redshifts of these two datasets separately. For one-step model, the performance of these three algorithms on photometric redshift estimation is compared with different training samples, and CatBoost is superior to XGBoost and Random forest. For two-step model, the performance of these three algorithms on the classification of low- and high-redshift subsamples are compared, and CatBoost still shows the best performance. Therefore CatBoost is regard as the core algorithm of classification and regression in two-step model. By contrast with one-step model, two-step model is optimal when predicting photometric redshift of quasars, especially for high redshift quasars. Finally the two models are applied to predict photometric redshifts of all quasar candidates of BASS DR3. The number of high redshift quasar candidates is 3938 (redshift $\ge 3.5$) and 121 (redshift $\ge 4.5$) by two-step model. The predicted result will be helpful for quasar research and follow up observation of high redshift quasars.
\end{abstract}


\section{Introduction} \label{sec:intro}
Redshift is an important feature of celestial objects and reflects the distance between celestial objects and the earth. By the redshift, the distance between celestial objects and the earth can be measured, which is of great significance for the research about spatial position, formation and evolution, and luminosity function of celestial objects. In general, the redshifts of celestial objects can be got more accurately through the spectra of celestial objects. However, the spectral observation of large-scale celestial objects is a time-consuming task, and especially for faint sources, it is almost impossible to get their spectra at present. With the construction and operation of many large sky survey observation equipments, a great amount of multi-band photometric data are obtained. Therefore, the study of redshift estimation of celestial objects through photometric data has great significance.
Although we cannot directly get the redshifts of celestial objects from photometric data, the relationship between photometric data and redshifts of the celestial objects can be reflected through a specific algorithm. \citet{Baum57} and \citet{Koo1985} proposed methods to estimate the redshifts of galaxies based on photometric data. Their results show that the redshifts of celestial objects can be measured well through multi-band photometric data.

In recent years, machine learning methods have been widely used in photometric redshift estimation. For instance, $k$-Nearest Neighbors ($k$NN; \citealt{ball07, zhang13}), Gaussian process regression \citep{way06, way09, bon10}, kernel regression \citep{wang07}, Self-Organizing-Map (SOM; \citealt{way12, car14}), Support Vector Machine (SVM; \citealt{jon17, sch17, Jinx2019}), Random Forest (RF; \citealt{car10, sch17}), Artificial Neural Networks (ANNs; \citealt{fir03, zhang09, yec10, cav12, bre13, cav17}), XGBoost \citep{Jinx2019} and Deep learning (DL; \citealt{Curran2021}) etc. Moreover researchers continuously try to develop new algorithms, improve old methods, or make innovations in algorithm applications.	
\citet{hoy16} applied Deep Neural Networks to estimate photometric redshifts of galaxies by using the full galaxy image in each measured band.
\citet{lei17} presented a new method to infer photometric redshifts in deep galaxy and quasar surveys, which combined the advantages of both machine learning methods and template fitting methods by building template spectral energy distributions (SEDs) directly from the spectroscopic training data.
\citet{zhang2019} put forward a new strategy for photometric redshift estimation of quasars. \citet{Han2021} devised a new approach GeneticKNN based on $k$NN and genetic algorithm for photometric redshift estimation of quasars.
	
Although a large number of algorithms have been used in this field, there is still large room for improvement. Moreover, due to the continuous increase in the amount of photometric data, it is necessary to speed up training and predicting while improving accuracy. In this paper, we explore three methods (CatBoost, XGBoost and Random Forest) to estimate photometric redshifts of quasars and then compare two schemes (one- and two-step models) for photometric redshift estimation of quasars. The sample used for this issue is described in Section 2. Then the adopted methods are briefly introduced in Section 3. Based on the samples, the different schemes for photometric redshift estimation of quasars by CatBoost, XGBoost and Random Forest are depicted in detail and compared in Section 4. The introduction and application of the two-step model is presented in Section 5. Finally we summarize the results of this paper in Section 6.

\section{Data} \label{sec:data}
The Beijing-Arizona Sky Survey (BASS; \citealt{Huzou2017a,Huzou2017b}) and MOSAIC z-band Legacy Survey (MzLS; \citealt{Slav2016}) are optical imaging surveys to provide galaxy and quasar targets for follow-up observation by the Dark Energy Spectroscopic Instrument (DESI; DESI Collaboration et~al. 2016). They survey the northern Galactic cap at ${\delta} >$ 30$^{\circ}$ and cover about 5400 deg$^2$. The BASS DR3 was released in 2019, which contains the data from all BASS and MzLS observations from 2015 January to 2019 March \citep{HuZou2019}. The DR3 includes single-epoch photometric catalogue and co-added photometric catalogue. In this paper, we used co-added photometric catalogue from BASS DR3, which can be download from https://nadc.china-vo.org/data/data/bassdr3coadd/f.

The Sloan Digital Sky Survey (SDSS; \citealt{york00}) has been conducting sky survey for about 20 years, acquiring a large amount of spectral and photometric data. The DR16 quasar catalogue (DR16Q) from SDSS includes 750 414 quasars \citep{Blan17}.

The Large Sky Area Multi-object Fiber Spectroscopic Telescope (LAMOST; \citealt{Cui12,Luo15}) may observe 4000 spectra in an observation to a limiting magnitude as faint as $r=19$ at the resolution $R=1800$. The first phase sky survey in five years has been finished. The number of quasars in the fifth data release (DR5, http://dr5.lamost.org/) adds up to 52 453 quasars.

The Wide-field Infrared Survey Explorer (WISE; \citealt{Wright10}) is an all sky survey project in mid-infrared band. On the basis of the WISE work, the AllWISE program has obtained better data than WISE on the respect of photometric sensitivity and accuracy as well as astrometric precision.

For simplicity, we use the released catalogue in \citealt{Li2021}, which provides optical information from BASS DR3 and infrared information from ALLWISE. From this catalogue, we select all possible quasar candidates when one of $Class\_b$, $Class\_bi$, $Class\_m$ and $Class\_mi$ is zero, the total number is 26 200 778. These selected sources are cross-matched with SDSS DR16Q and LAMOST DR5 quasars in 2 arcsec radius respectively. Then we obtain known samples BS\_W and BL\_W with spectral redshifts respectively from SDSS DR16Q and LAMOST DR5. For the known samples, we extinction-correct all photometries and use AB magnitudes referring to the work \citep{sch17}. The parameters about known samples are extracted from BASS, ALLWISE, SDSS and LAMOST databases, as described in Table~1. The spectroscopic redshift distributions of the known samples are shown in Figure~1.

\begin{table*}
	\begin{center}
    		\caption{Parameters in the known samples}
    		\begin{tabular}{rlll}
    			\hline
    			Parameters&Definition  &Catalogue& Waveband\\
    			\hline
    			id &Source ID      &BASS   &\\
    			ra &Right ascension in decimal degrees &BASS &\\
    			dec &Declination in decimal degrees    &BASS &\\
    			$gKronMag$ &Kron magnitude in $g$ band  &BASS &Optical band\\
    			$rKronMag$ &Kron magnitude in $r$ band  &BASS &Optical band\\
    			$zKronMag$ &Kron magnitude in $z$ band  &BASS &Optical band\\
    			$gPSFMag$   &PSF magnitude in $g$ band  &BASS &Optical band\\
    			$rPSFMag$   &PSF magnitude in $r$ band  &BASS &Optical band\\
    			$zPSFMag$   &PSF magnitude in $z$ band  &BASS  &Optical band\\
    			$g$     & extinction-corrected PSF magnitude in $g$ band  &BASS & Optical band\\
    			$r$     & extinction-corrected PSF magnitude in $r$ band  &BASS & Optical band\\
    			$z$    & extinction-corrected PSF magnitude in $z$ band  &BASS & Optical band\\
    			$W1mag$   &$W1$ magnitude  &ALLWISE &Infrared band\\
    			$W2mag$   &$W2$ magnitude  &ALLWISE &Infrared band\\
    			$W1$   & extinction-corrected $W1$ magnitude &ALLWISE& Infrared band\\
    			$W2$   & extinction-corrected $W2$ magnitude &ALLWISE& Infrared band\\
    			$Redshift$ &Spectral redshift  &SDSS, LAMOST &\\
    			\hline
    		\end{tabular}
    		\bigskip
    	\end{center}
    \end{table*}

    	\begin{figure*}
    		\centering
    		\includegraphics[bb=84 239 531 558,width=8cm,height=6.5cm]{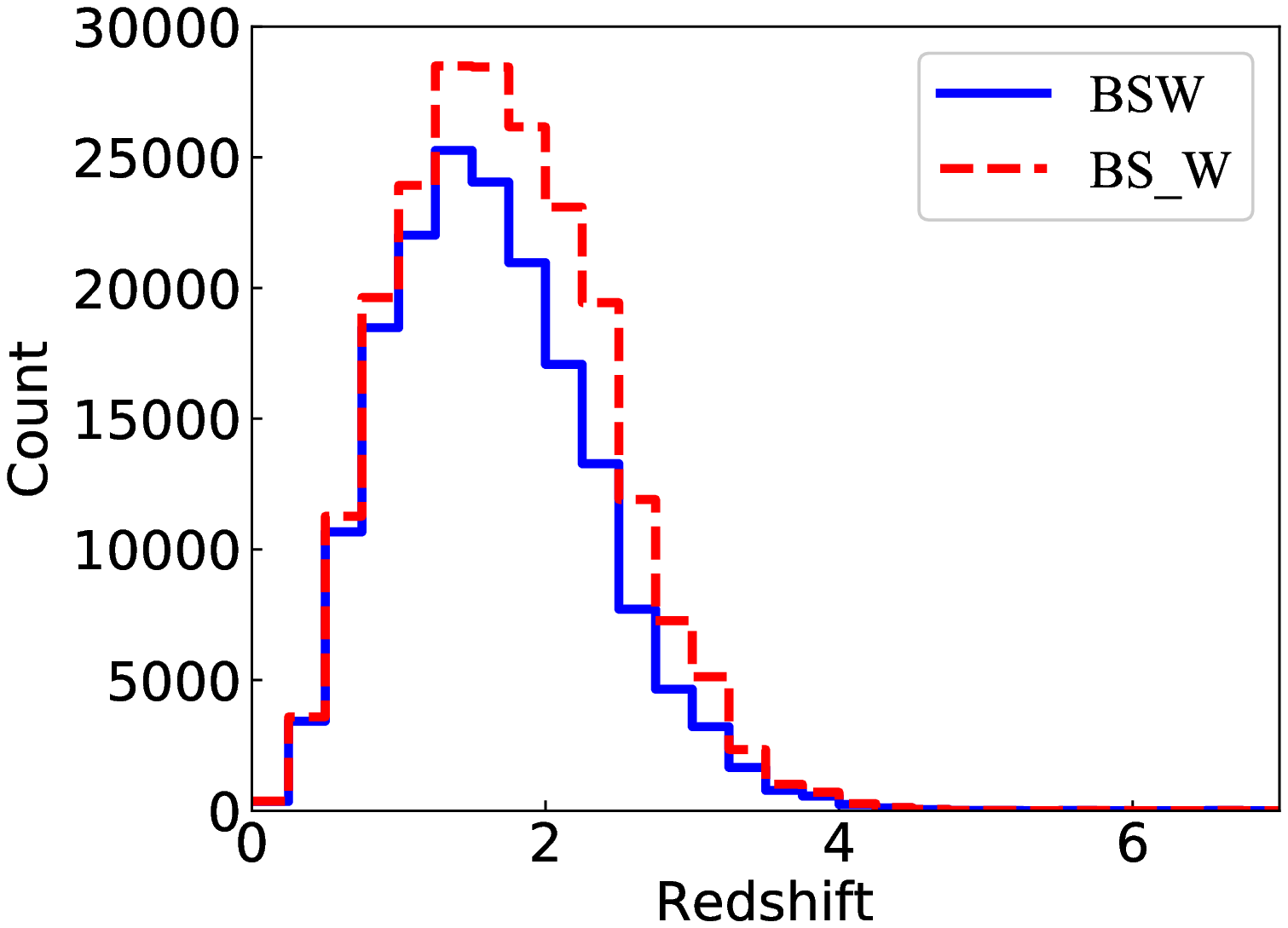}
    		\includegraphics[bb=84 239 531 558,width=8cm,height=6.5cm]{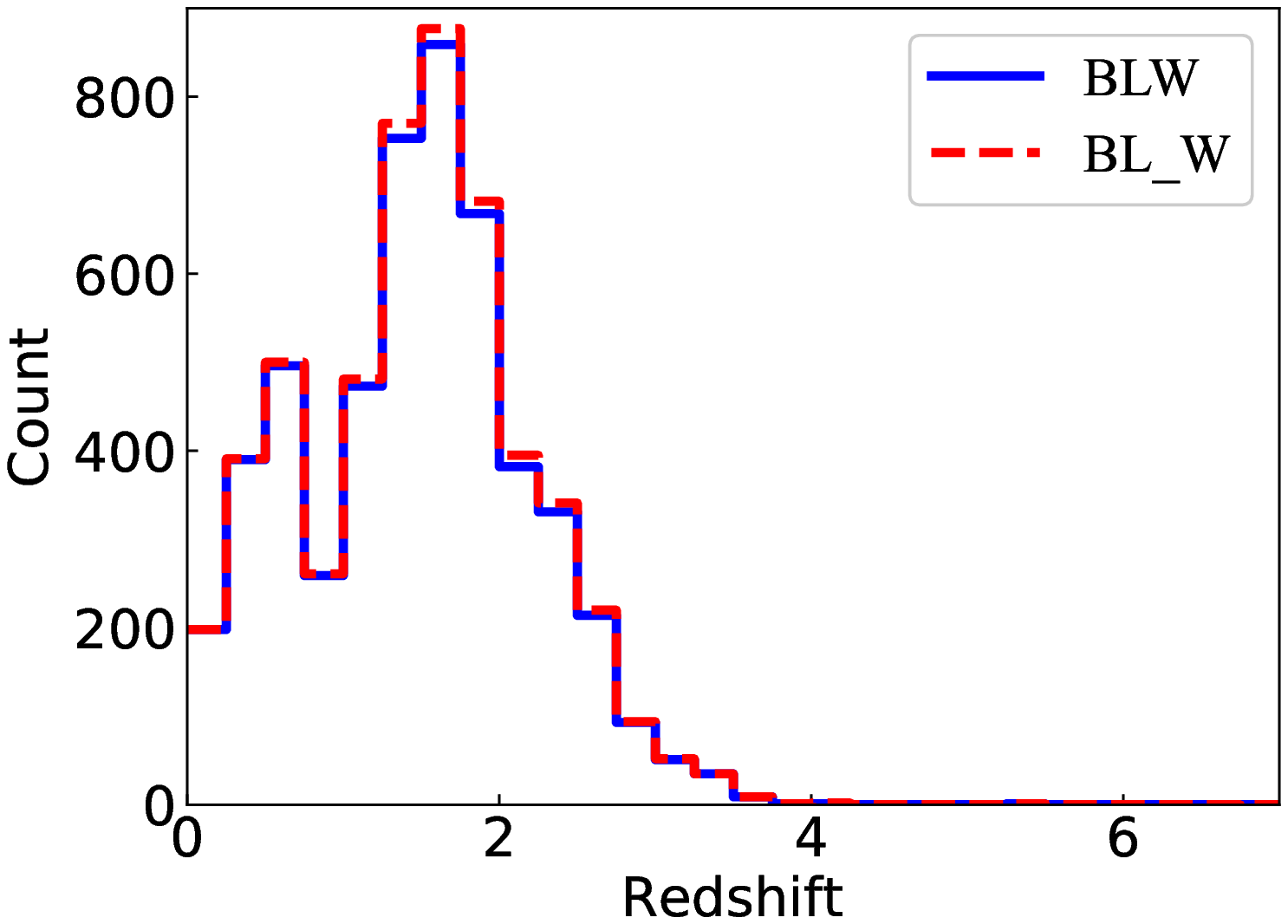}
    		\caption{Left panel: the distribution of spectroscopic redshifts for known samples BSW (blue line) \& BS\_W (red dash line); right panel: the distribution of spectroscopic redshifts for known samples BLW (blue line) \& BL\_W (red dash line). }	
    		\label{fig1}
    	\end{figure*}

Then we split the sample BS\_W into two datasets BSO and BSW according to whether both $W1mag$ and $W2mag$ are $NULL$ or not. When both $W1mag$ and $W2mag$ are $NULL$, this source belongs to the sample BSO, otherwise, it belongs to the sample BSW. Correspondingly, the sample BL\_W is divided into two datasets BLO and BLW. In this paper, the samples BS\_W and BSW are used as training set and test set, BL\_W and BLW are used as external test set.

\section{Method} \label{sec:method}
\subsection{CatBoost}
Gradient Boosted Decision Trees (GBDT, \citealt{Friedman2001}) are a powerful tool for classification and regression tasks. CatBoost is developed by Yandex researchers and engineers \citep{catboost2018}, and a high-performance open source algorithm. It is a member of the family of GBDT machine learning ensemble techniques, which applies boosting method to build strong classifiers by means of learning multiple weak classifiers or regressors. It uses Oblivious Decision Trees (ODT) to build decision trees, which are full binary trees, furthermore, all non-leaf nodes of ODT will have the same splitting criteria. This design is helpful to speed up the score and avoid overfitting. It supports categorical features and text features in data without additional preprocessing, and uses the ordered target statistics method for encoding new features. Besides, it is easy to get good result with the default set of model parameters. Thus, a lot of time for tuning its model parameters will be saved. It also has fast and scalable GPU version fit for handling big data. Compared to GBDT, it has two innovations: ordered target statistics and ordered boosting.

\subsection{XGBoost}
XGBoost \citep{Chen2016} is an excellent ensemble learning algorithm. Compared with other ensemble models, it can improve the model's robustness by introducing regular terms and column sampling; and when each tree selects the split point, a parallelization strategy will be adopted to improve the model's running speed. Besides, XGBoost can overcome the limitations of computational speed and accuracy to a certain extent, require less training and prediction time, and support various objective functions, when performing classification and regression tasks.

\subsection{Random Forest}
Random Forest (RF, \citealt{Bre01}) is based on bagging models built using the decision tree method. RF uses bootstrap resampling technology to randomly select $K$ subsamples from the original training sample with replacement to generate a new training sample set, and build $K$ classification trees to form a random forest. Each tree in the forest has the same distribution, and the classification error depends on the classification ability of each tree and the correlation between them. Feature selection uses a random method to split each node, and then compares the errors generated in different situations. Therefore RF uses the average to improve the predictive accuracy and control over-fitting. In the scikit-learn implementation of RF combines classifiers by averaging their probabilistic prediction, instead of letting each classifier vote for a single class.

XGBoost and Random Forest methods have wide applications in astronomy, such as classification of unknown source \citep{Mirabal2016}, quasar candidate selection \citep{Jinx2019}, photometric redshift estimation \citep{zhang2019} etc. CatBoost has various applications in different fields, such as finance, investment, petroleum and medicine. In this paper, we use XGBoost, CatBoost and Random Forest as supervised learning algorithms to build regressors for photometric redshift estimation of different quasar samples with different features. We also compare the performance of these three algorithms when classifying quasar sample into low- and high-redshift subsamples. CatBoost, XGboost and Random Forest python packages are provided by scikit-learn \citep{scikit-learn} and all computing runs in the cloud computing environment of National Astronomical Data Centre (NADC) \citep{Li2017}.

\section{Photometric redshift estimation} \label{sec:performance}
\subsection{Regression Metrics}
The performance evaluation of different algorithms about photometric redshift estimation depends on different metrics, such as the residual between the spectroscopic and photometric redshifts, $\Delta \rm z=\rm z_{\rm spec} - \rm z_{\rm photo}$, the mean absolute error (${\rm MAE}$) and the mean squared error (${\rm MSE}$). They are defined as follows:
\begin{equation}
	MAE = \frac{1}{n}\sum_{i=0}^{n-1} | z_{i}-\widehat{z}_{i} | \\
\end{equation}

\begin{equation}
   MSE =\frac{1}{n}\sum_{i=0}^{n-1} ( z_{i}-\widehat{z}_{i} )^2 \\
   \end{equation}
where $z_{i}$ is the true redshift, $\widehat{z_{i}}$ is the predicted redshift value and $n$ is the sample size.

The fraction of test sample that satisfies $|\rm \Delta{z}| <e$ is usually used to evaluate the redshift estimation, where $e$ is a given residual threshold (\citealt{sch17} and references therein). In reality, the redshift normalized residual ($\rm \Delta{z(norm)}$) is often adopted, and we use $e=0.3$.
\begin{equation}
\rm{{\Delta}z(norm)} = \frac{z_{\rm spec} - z_{\rm phot}}{ 1 + z_{\rm spec}}  \\
\end{equation}

\begin{equation}
\rm{\delta_{0.3}} = \frac{\rm N_{\left|{\Delta}z(norm) \right| < 0.3}}{\rm N_{total}}
\end{equation}

Then, we use five additional metrics defined below as a reference for performance evaluation of different machine learning methods: $\mathrm R^2$, bias (the average separation between prediction and true values), the standard deviation between the photometric redshifts and the spectroscopic redshifts, the normalized median absolute deviation ($\sigma_\mathrm{NMAD}$) and the outlier fraction (O) \citep{Ben2021,Curran2021}.
\begin{equation}
\mathrm{R^2}=1 -  \frac{\sum_{i=1}^{n} | (z_{i}-\widehat{z}_{i} )^2 }{\sum_{i=1}^{n} ( z_{i}-\overline{z} )^2 } \\
\end{equation}

\begin{equation}
\mathrm{Bias}=< z_{\rm spec} - z_{\rm phot}> \\
\end{equation}

\begin{equation}
\sigma_{{\Delta}\rm z} =\sqrt{\frac{1}{n}\sum_{i=0}^{n-1} (\rm  {\Delta}z )^2 }\\
\end{equation}

\begin{equation}
\sigma_{\rm  NMAD} = 1.48 \times \rm{median} \left| \frac{z_{\rm spec} - z_{\rm phot}}{ 1 + z_{\rm spec}} \right|\\
\end{equation}

\begin{equation}
Outlier\,\, fraction\,\, (O) = \frac{\rm N_{\left|{\Delta}z(norm) \right| > 0.15}}{\rm N_{total}}
\end{equation}

\subsection{Feature selection}
When handling high dimensional data, feature selection is the key factor influencing the performance of a machine learning algorithm. Feature selection not only reduces the dimension of data and rules out unimportant features, but also contributes to improve the accuracy of an algorithm. During data preprocessing, we need to obtain optimal features firstly, and adopt two steps for feature selection.

According to the parameters listed in Table~1, we define optical and infrared features. All optical features include $KronMag - PSFMag$ of g,r,z (namely ${\Delta}g$, ${\Delta}r$, ${\Delta}z$, respectively), $g$, $r$, $z$, $g-r$, $r-z$, $g-z$, and all infrared features are $W1$, $W2$, $g-W1$, $r-W1$, $z-W1$, $g-W2$, $r-W2$, $z-W2$, $W1-W2$. For the samples BSW and BS\_W, they have all optical and infrared features. The difference between the samples BSW and BS\_W is that the sample BS\_W includes the samples BSW and BSO. Each source in the sample BSW contains both optical and infrared information while the sample BSO only includes optical information. The difference between the samples BLW and BL\_W is same. The quasars of BSW and BS\_W are from SDSS DR16Q, and the quasars of BLW and BL\_W are from LAMOST DR5. As for the samples BSO and BLO, only the features from optical band are used. The sources in the samples BS\_W and BL\_W have no match sources in ALLWISE and no infrared features, so the features of these sources related to infrared band are missing. In other words, the samples BS\_W and BL\_W contain missing values. We set all missing values to 0 when using Random Forest. XGBoost and CatBoost support missing values.

Firstly, all features of training samples are evaluated by Random Forest, XGBoost and CatBoost methods, which can give the importance score of each feature. The importance type ($importance\_type$) is $total\_gain$. We sort these features by the importance score. The feature importance for different samples by different methods is shown in Figure~2. Figure~2 indicates that the feature importance is closely related to samples and algorithms. Furthermore, from the feature importance rank, it is seen that the infrared information is very important to the redshift estimation of quasars.

\begin{figure*}
\centering
\subfigure[For the sample BSW with optical and infrared features]{
\includegraphics[height=5cm,width=5.5cm]{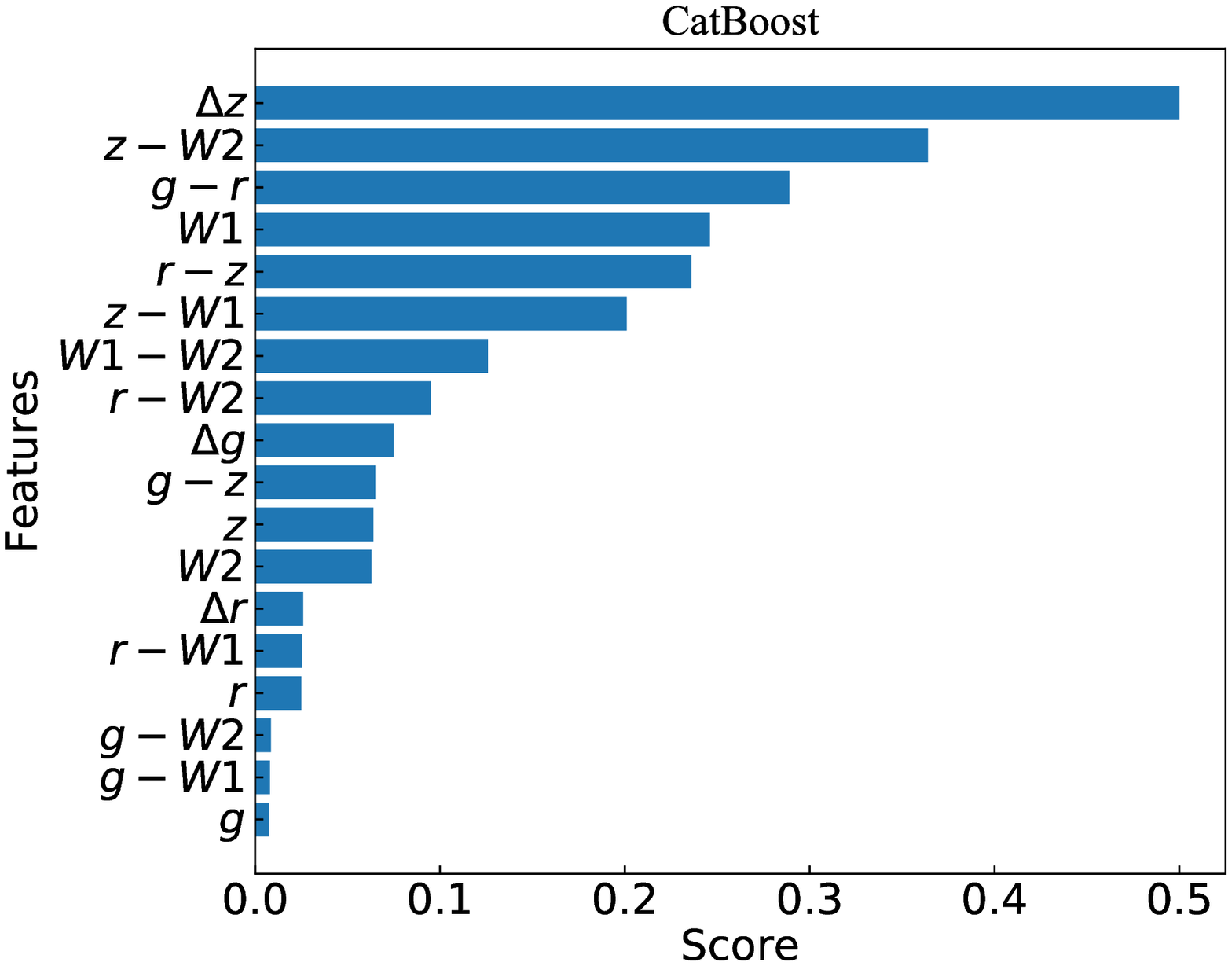}
\includegraphics[height=5cm,width=5.5cm]{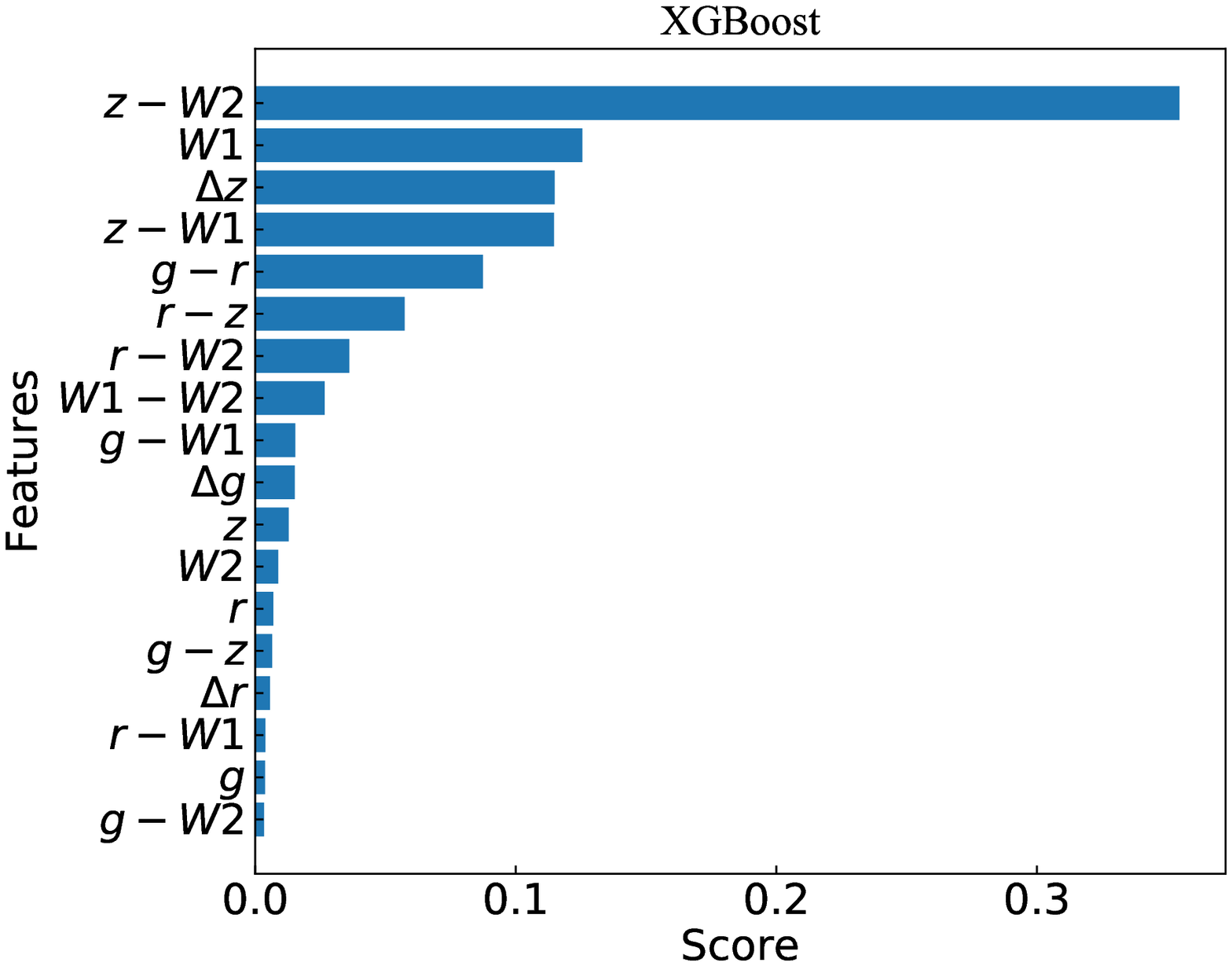}
\includegraphics[height=5cm,width=5.5cm]{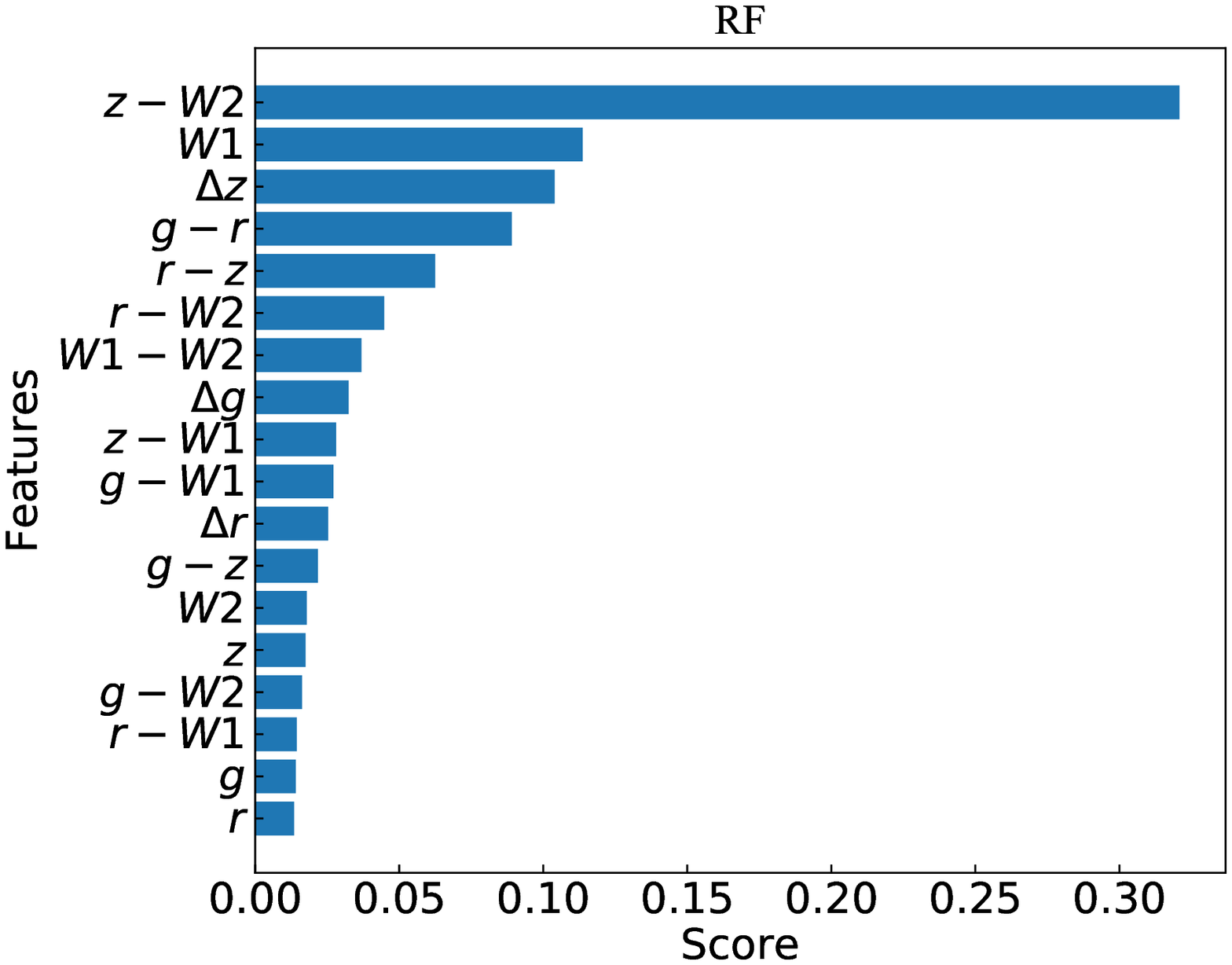}
}
\subfigure[For the sample BS\_W with optical and infrared features]{
\includegraphics[height=5cm,width=5.5cm]{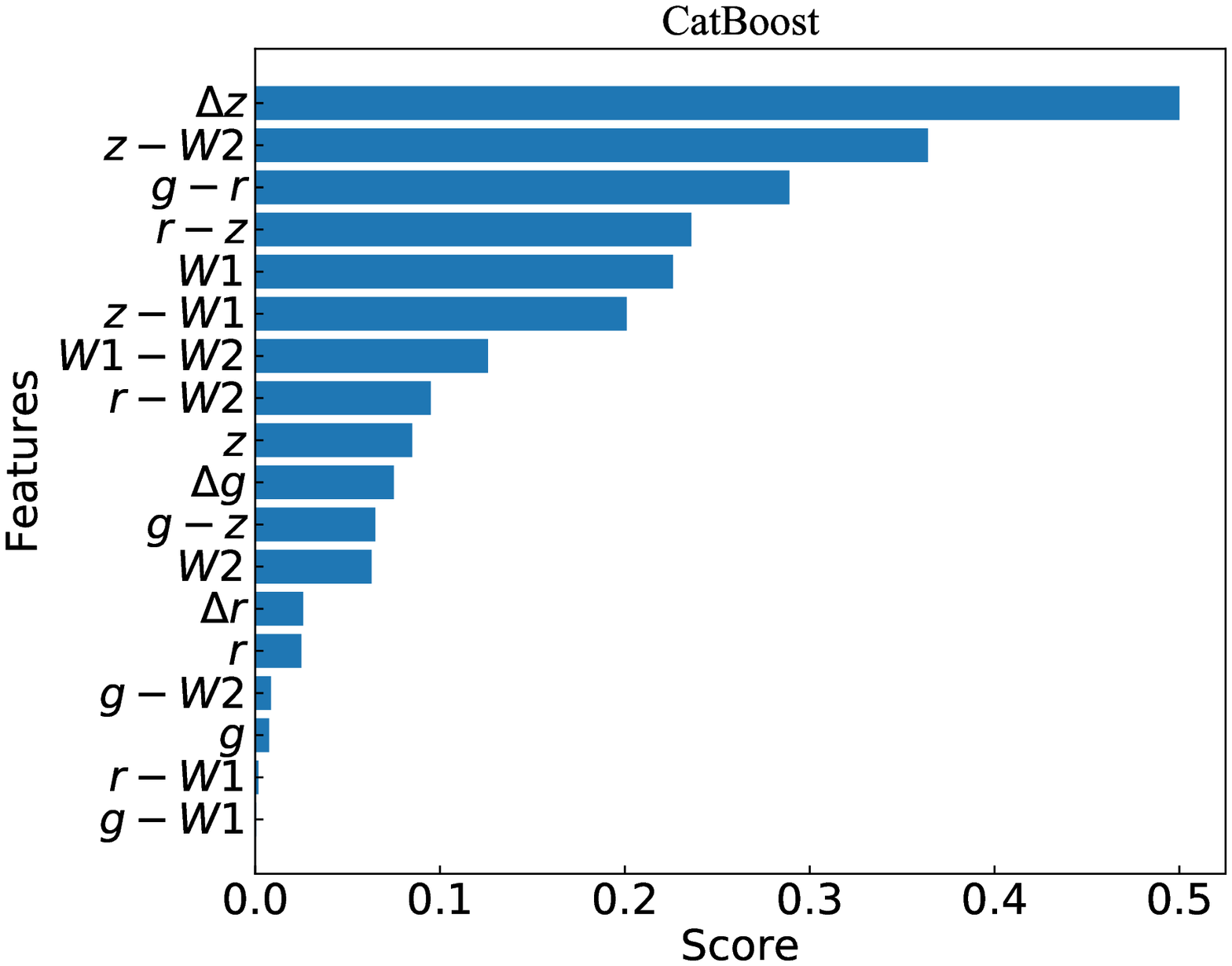}
\includegraphics[height=5cm,width=5.5cm]{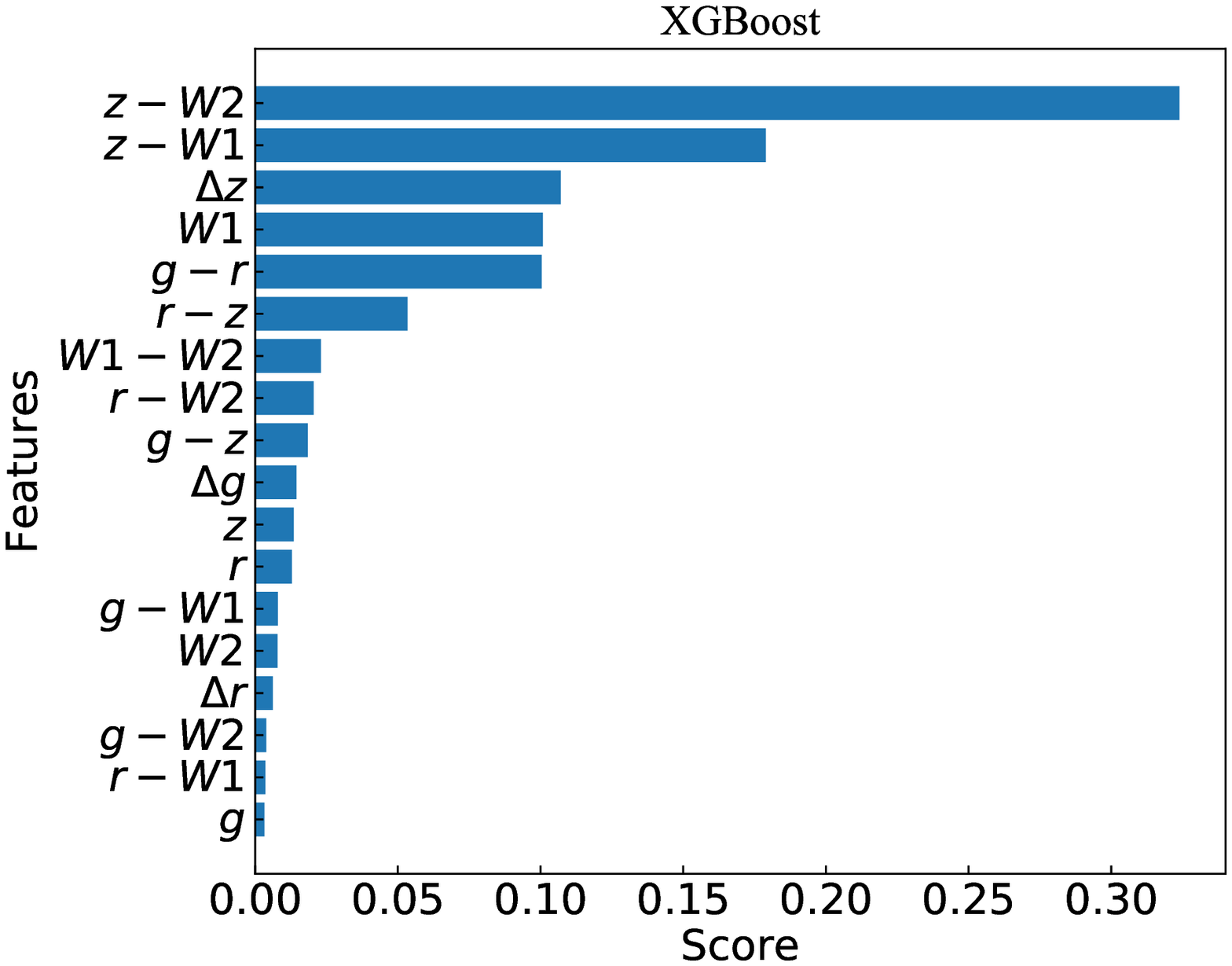}
\includegraphics[height=5cm,width=5.5cm]{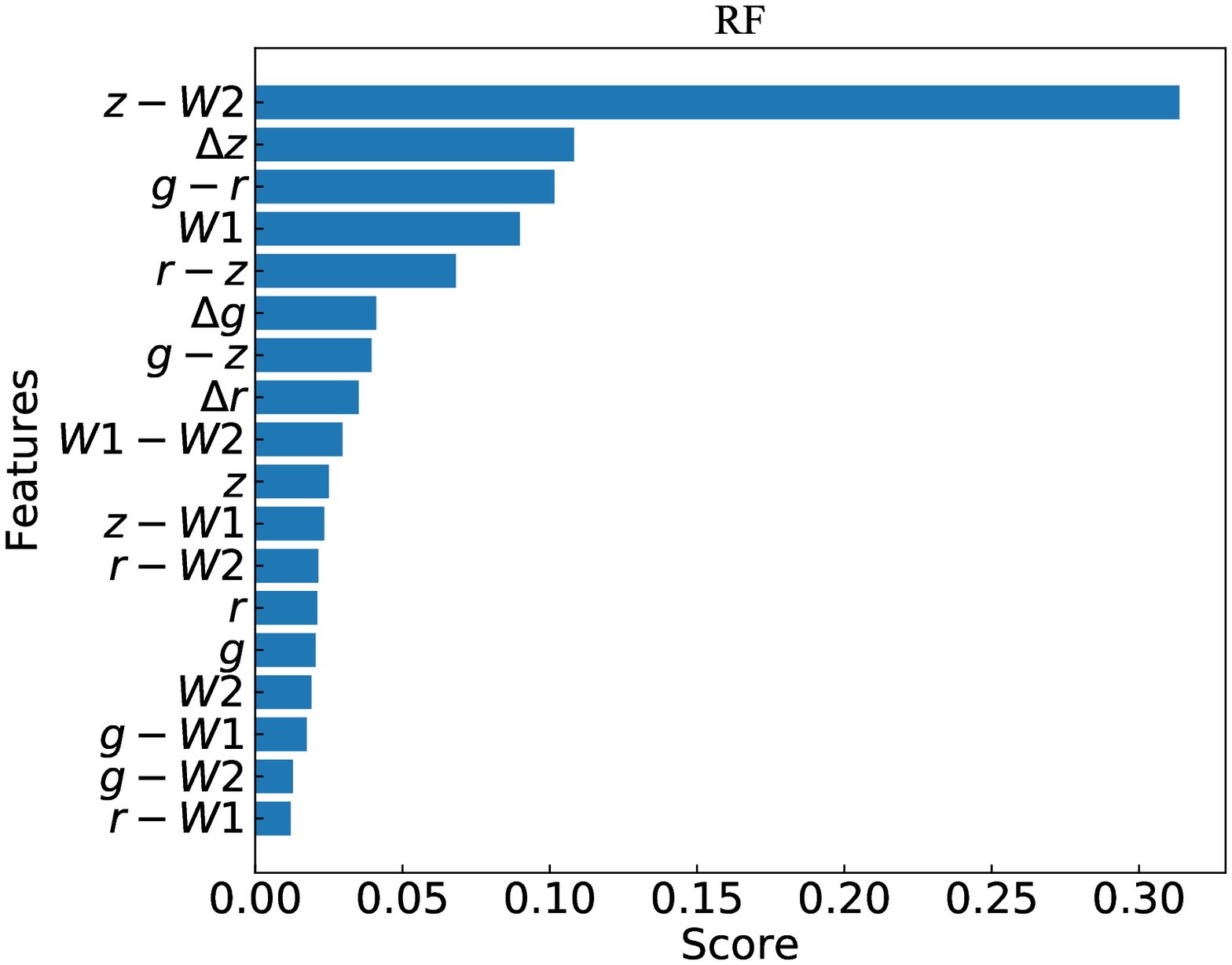}
}
\subfigure[For the sample BS\_W only with optical features]{
	\includegraphics[height=5cm,width=5.5cm]{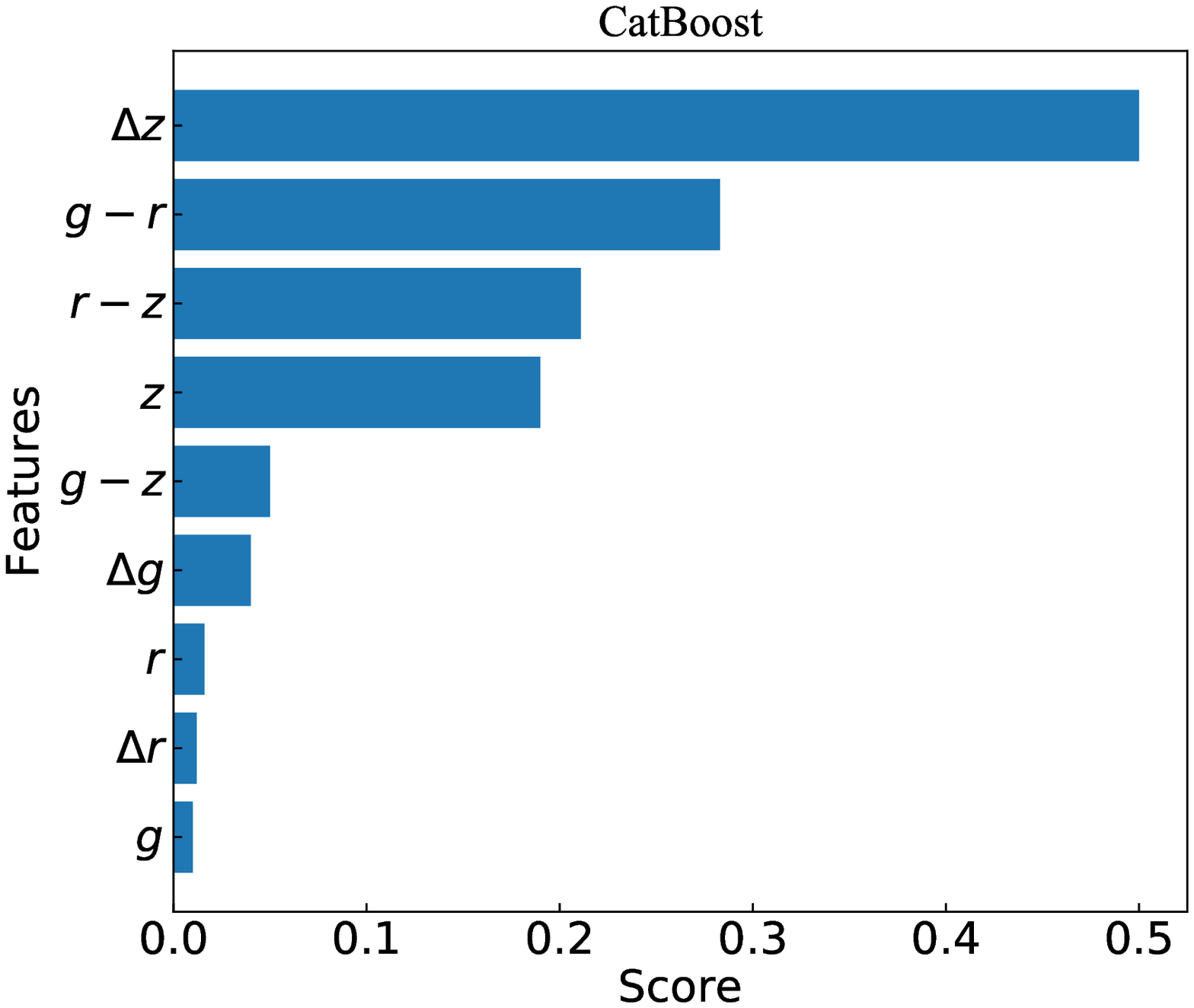}
	\includegraphics[height=5cm,width=5.5cm]{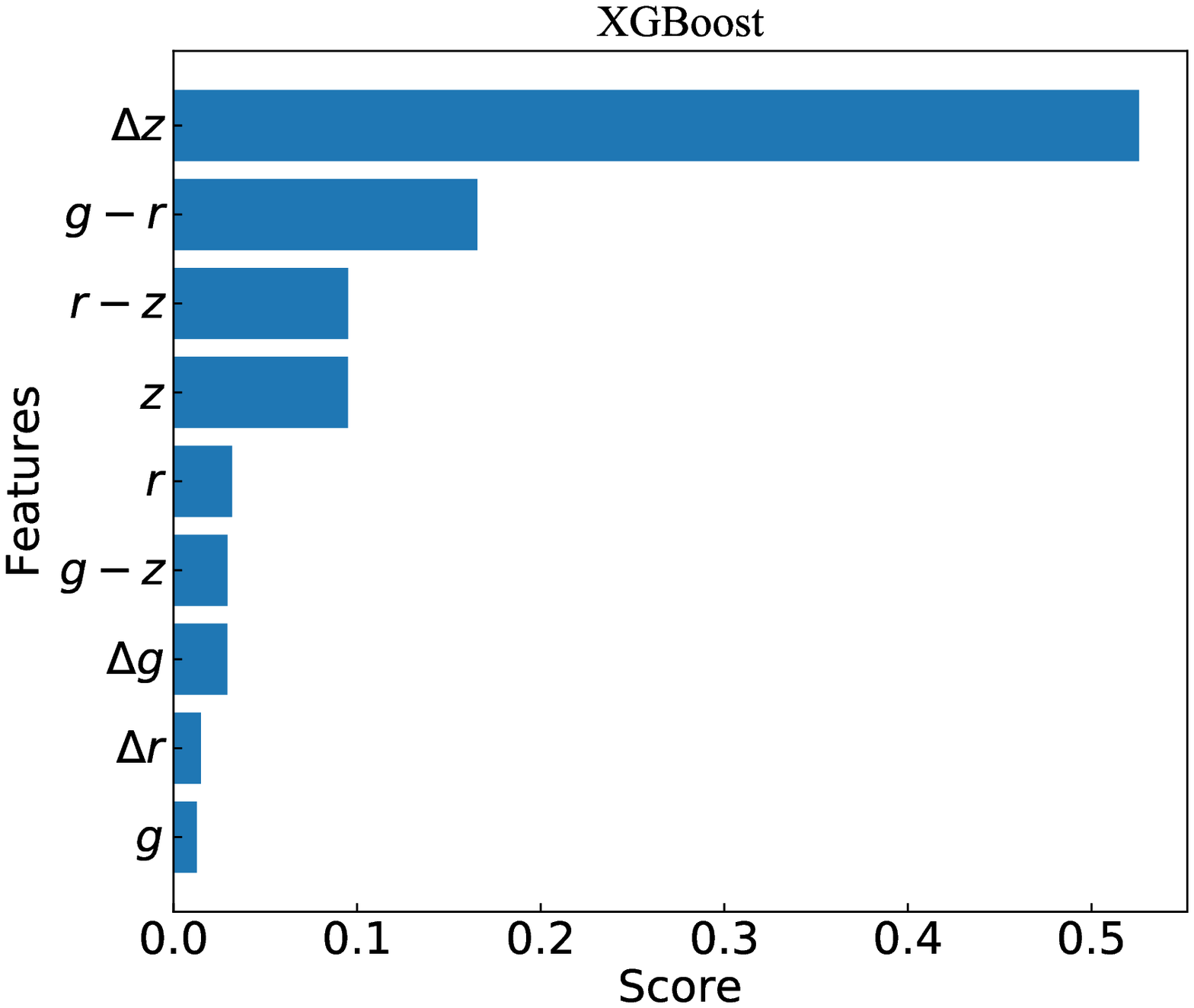}
	\includegraphics[height=5cm,width=5.5cm]{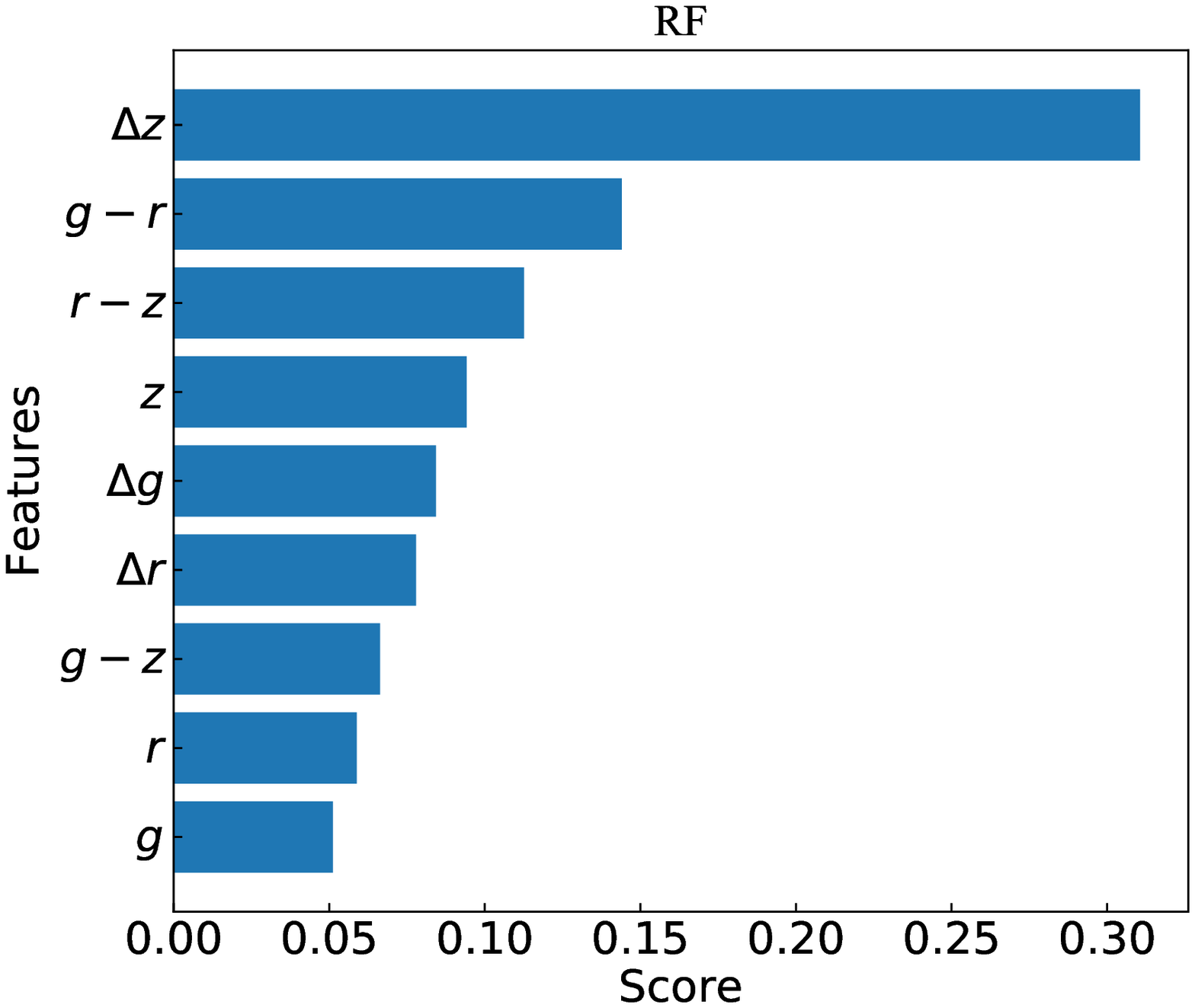}
}
\caption{The feature importance for different samples by CatBoost, XGBoost and Random Forest.}
\label{fig2}
\end{figure*}

Secondly, when a method is adopted, according to the above rank of features, we select the top four features as initial input pattern to train a model, then add one feature in turn to the input pattern for training and record all performance. $MSE$ with different input features for different methods is described in Figure~3. For simplicity, the best input pattern is adopted when $MSE$ achieves the minimum. As shown in the left panel of Figure~3, for the sample BSW, the best input pattern is $z-W2$, $W1$, ${\Delta}z$, $z-W1$, $g-r$, $r-z$, $r-W2$, $W1-W2$, $g-W1$, ${\Delta}g$, $z$ (11 features) by XGBoost; the best input pattern is ${\Delta}z$, $z-W2$, $g-r$, $W1$, $r-z$, $z-W1$, $W1-W2$, $r-W2$, ${\Delta}g$, $g-z$, $z$, $W2$, ${\Delta}r$ (13 features) by CatBoost, and the best input pattern is $z-W2$, $W1$, ${\Delta}z$, $g-r$, $r-z$, $r-W2$, $W1-W2$, ${\Delta}g$, $z-W1$, $g-W1$, ${\Delta}r$, $g-z$, $W2$, $z$, $g-W2$ (15 features) by Random Forest. As plotted in the middle panel of Figure~3, for the sample BS\_W, the optimal input pattern of XGBoost is $z-W2$, $z-W1$, ${\Delta}z$, $W1$, $g-r$, $r-z$, $W1-W2$, $r-W2$, $g-z$, ${\Delta}g$, $z$, $r$, $g-W1$, $W2$, ${\Delta}r$, $g-W2$ (16 features); the optimal input pattern of CatBoost is ${\Delta}z$, $z-W2$, $g-r$, $r-z$, $W1$, $z-W1$, $W1-W2$, $r-W2$, $z$, ${\Delta}g$, $g-z$, $W2$, ${\Delta}r$, $r$, $g-W2$ (15 features), and the optimal input pattern of Random Forest is $z-W2$, ${\Delta}z$, $g-r$, $W1$, $r-z$, ${\Delta}g$, $g-z$, ${\Delta}r$, $W1-W2$, $z$, $z-W1$, $r-W2$ (12 features). As indicated in the right panel of Figure~3, for the sample BS\_W only with optical features, the optimal input patterns of the three methods keep all optical features (9 features) and retain their order. With the best input patterns for different samples by the three methods, we adopt the default hyper parameters of models and 5-fold cross validation to get the average values of $MSE$, $MAE$, bias, $\sigma_{\rm  NMAD}$, $\sigma_{{\Delta}\rm z}$, $R^2$, ${\delta}_{0.3}$, $Outlier\,\, fraction\,\, (O)$ and running $time$, which are shown in Table~2. Comparing all metrics in Table~2, CatBoost achieves better performance than XGBoost and Random Forest for most metrics. In general, the better performance with optical and infrared features is archived than only with optical features. Table~2 further confirms this fact.

    \begin{figure*}
    	\centering    	
    	\includegraphics[height=5cm,width=5.5cm]{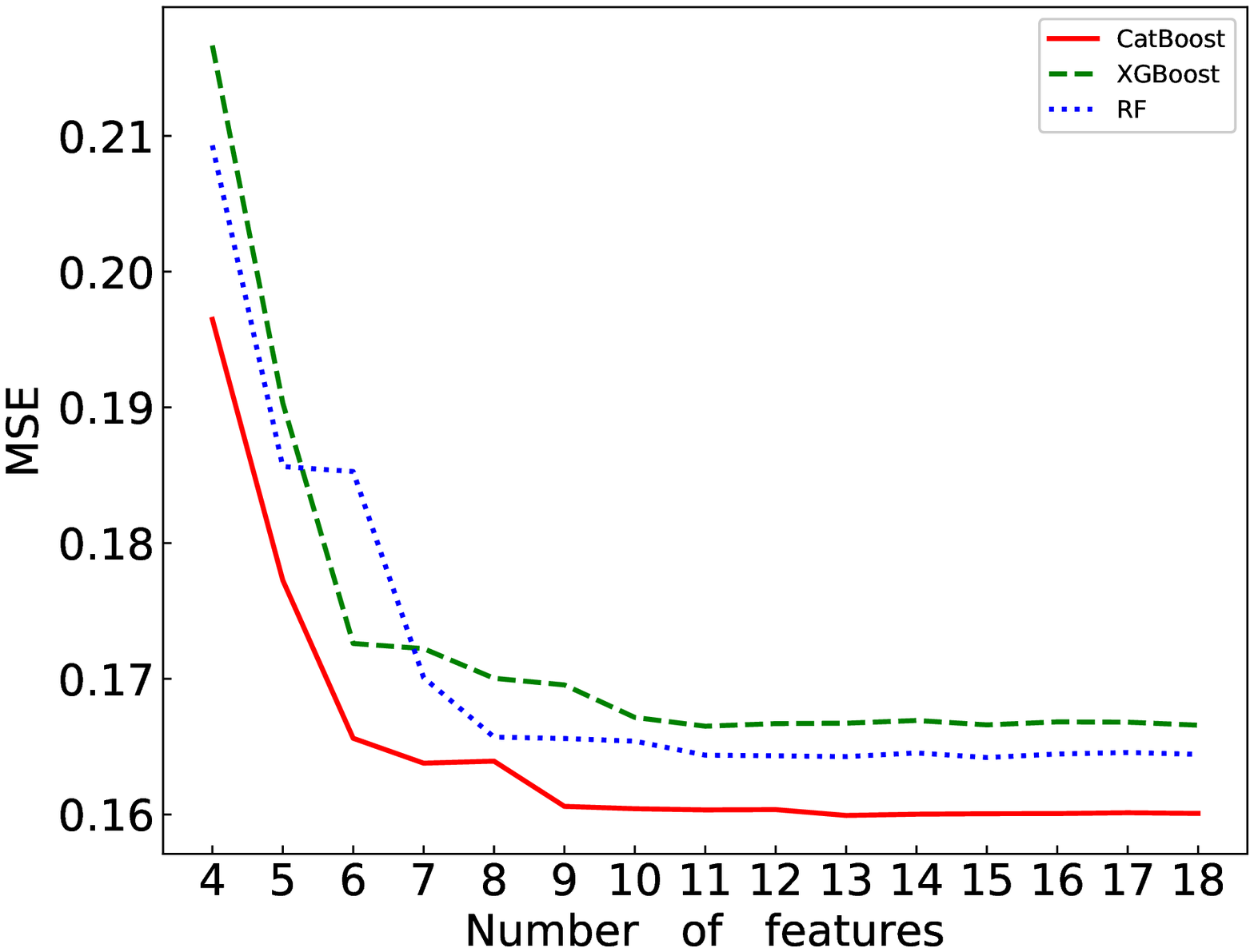}
    	\includegraphics[height=5cm,width=5.5cm]{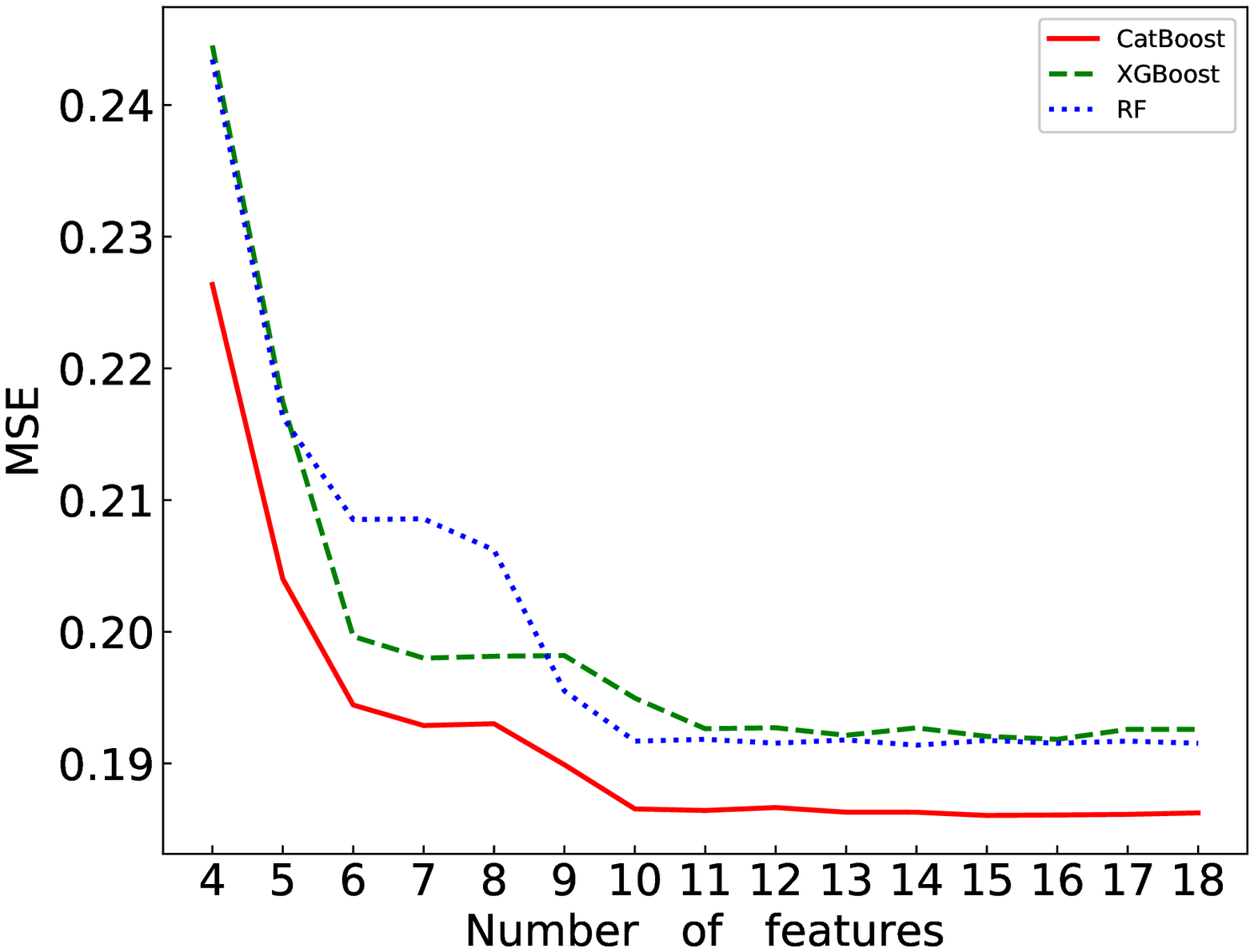}
    	\includegraphics[height=5cm,width=5.5cm]{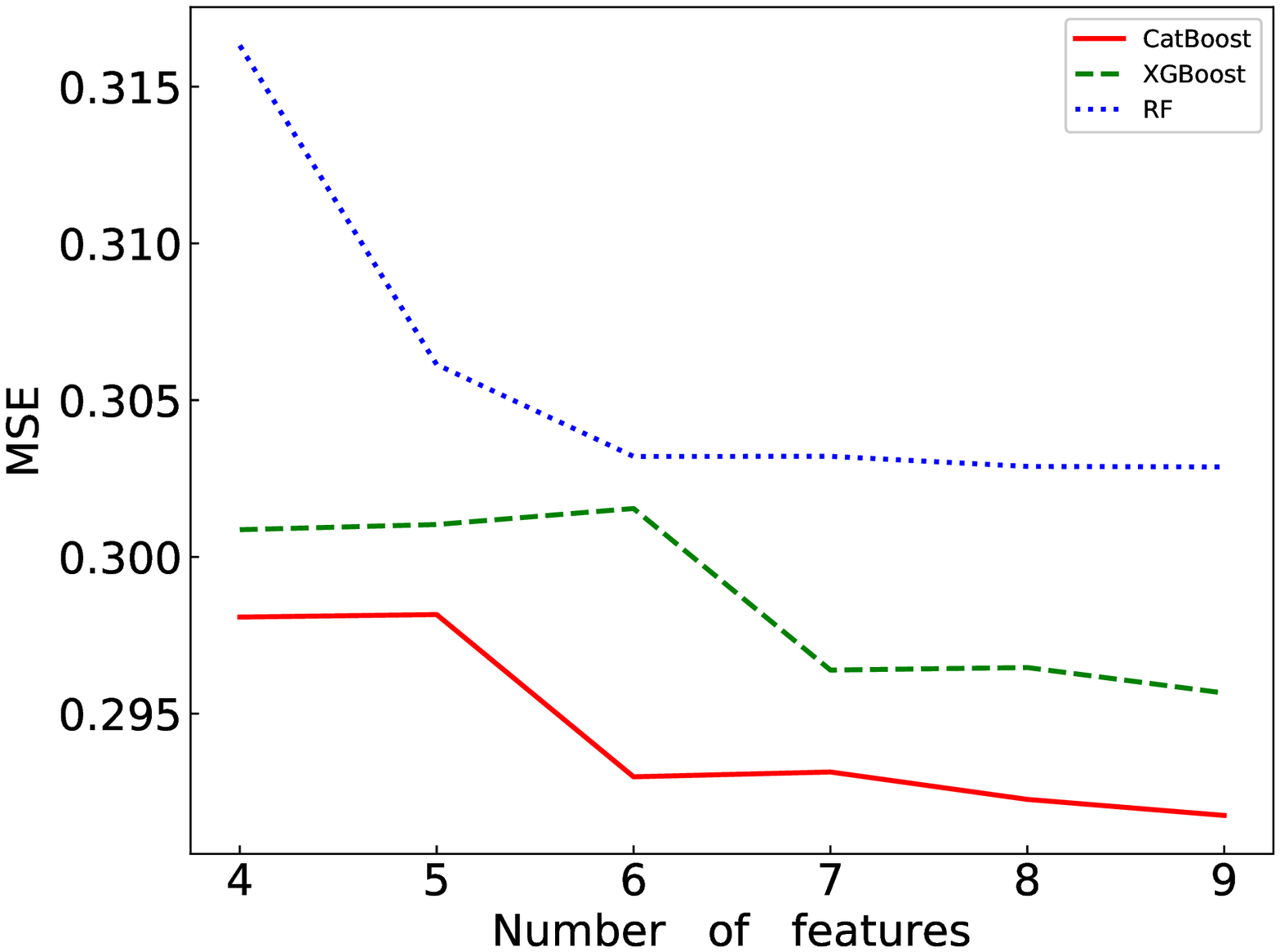}	    	
    	\caption{The performance of different methods with different input patterns for different samples. Left panel: for the sample BSW with optical and infrared information; middle panel: for the sample BS\_W with optical and infrared information; right panel: for the sample BS\_W only with optical information. }
    	\label{fig3}
    \end{figure*}

      \begin{table*}
      	\begin{center}
      		\caption[]{The performance of photometric redshift estimation with optimal input features for each method with default model parameters\label{tab:fsel}}
      		\begin{tabular}{rccccccccccc}
      			\hline
      			Sample & Input pattern   &  Method  &MSE &MAE& $\mathrm{Bias}$ & ${\sigma}_{\rm NMAD}$& ${\sigma}_{{\Delta}\rm z}$&$R^2$&${\delta}_{0.3}$(\%)&$O$(\%)&Time(s)\\
      			\hline
      			BSW   &	Pattern I   &XGBoost & 0.1666&0.2765&4.0 $\times 10^{-5}$&0.1070&0.4082&0.6493&93.39&22.04&3\\
      			BSW   &	Pattern II  &CatBoost & 0.1600 &0.2702&6.0 $\times 10^{-5}$&0.1036&0.4000&0.6632&93.61&21.32&10 \\
      			BSW   &	Pattern III &RF &0.1645 &0.2700&2.2 $\times 10^{-3}$&0.1008&0.4056&0.6537&93.35&21.01&442 \\
      			
      			BS\_W   &	Pattern IV &XGBoost &0.1925&0.3064&-7.6$\times 10^{-6}$&0.1171&0.4387&0.6071&92.37&25.22&6\\
      			BS\_W   &	Pattern V  &CatBoost &0.1867&0.3012&-4.8$\times 10^{-6}$&0.1146&0.4321&0.6189&92.56&24.48&12 \\
      			BS\_W   &	Pattern VI &RF &0.1924 &0.3009&-2.0$\times 10^{-5}$&0.1107&0.4387&0.6074&92.32&24.03&447 \\
      			
      			BS\_W  &	Pattern VII  &XGBoost &0.2959 &0.4009& -1.0 $\times 10^{-4}$&0.1657&0.5439&0.4008&87.27&37.40&4 \\
      			BS\_W  &	Pattern VII  &CatBoost &0.2917 &0.3988&-9.0 $\times 10^{-5}$&0.1651&0.5403&0.4087&87.39&37.74&12 \\
      			BS\_W  &	Pattern VII  &RF&0.3032 &0.4045&8.7 $\times 10^{-3}$&0.1648&0.5507&0.3858&86.52&37.93&200 \\
      			\hline
      			
      			\multicolumn{12}{l}{\tiny{$^a$ Pattern I represents $z - W2$, $W1$, ${\Delta}z$, $z - W1$, $g - r$, $r - z$, $r - W2$, $W1 - W2$, $g - W1$, ${\Delta}g$, $z$ (11 features).}} \\
      			\multicolumn{12}{l}{\tiny{$^b$ Pattern II represents ${\Delta}z$, $z - W2$, $g - r$, $W1$, $r - z$, $z - W1$, $W1 - W2$, $r - W2$, ${\Delta}g$, $g - z$, $z$, $W2$, ${\Delta}r$ (13 features).}}\\
      			\multicolumn{12}{l}{\tiny{$^c$ Pattern III represents $z - W2$, $W1$, ${\Delta}z$, $g - r$, $r - z$, $r - W2$, $W1 - W2$, ${\Delta}g$, $z - W1$, $g - W1$, ${\Delta}r$, $g - z$, $W2$, $z$, $g - W2$ (15 features).}}\\
      			\multicolumn{12}{l}{\tiny{$^d$ Pattern IV represents $z - W2$, $z - W1$, ${\Delta}z$, $W1$, $g - r$,  $r - z$, $W1 - W2$, $r - W2$, $g - z$, ${\Delta}g$, $z$, $r$, $g - W1$, $W2$, ${\Delta}r$, $g - W2$ (16 features).}}\\
      			\multicolumn{12}{l}{\tiny{$^e$ Pattern V represents ${\Delta}z$, $z - W2$, $g - r$, $r - z$, $W1$, $z - W1$, $W1 - W2$, $r - W2$, $z$, ${\Delta}g$, $g - z$, $W2$, ${\Delta}r$, $r$, $g - W2$ (15 features).}}\\
      			\multicolumn{12}{l}{\tiny{$^f$ Pattern VI represents $z - W2$, ${\Delta}z$, $g - r$, $W1$, $r - z$, ${\Delta}g$, $g - z$, ${\Delta}r$, $W1 - W2$, $z$, $z - W1$, $r - W2$ (12 features).}}\\
      			\multicolumn{12}{l}{\tiny{$^g$ Pattern VII represents ${\Delta}z$, ${\Delta}g$, ${\Delta}r$, $g - r$, $g - z$, $r - z$, $g$, $r$, $z$ (9 features).}}\\
      			\multicolumn{12}{l}{\tiny{$^h$ Pattern I, II, III, IV, V, VI, VII represents the same meaning in the following of this paper.}}\\
      			\hline
      		\end{tabular}
      	\end{center}
      \end{table*}

\subsection{Model parameter optimization}
Since the optimal input patterns have been set, the next task is to determine the hyper parameters of models. Model parameter optimization is a very complex task. In order to reduce computing scale, we only choose some main hyper parameters for each method. For XGBoost, the important model parameters contain the maximum depth of individual trees (max\_depth) and the number of weak estimators (n\_estimators); for CatBoost, the key model parameters are the maximum depth of individual tree (depth) and the maximum number of trees (iterations); for Random Forest, the main model parameters are the maximum depth of individual trees (max\_depth) and the number of trees in the forest (n\_estimators). We adopt the grid search method to get the optimal model parameters and 5-fold cross validation to get the average values of $MSE$, $MAE$, $Bias$, ${\sigma}_{\rm NMAD}$, ${\sigma}_{{\Delta}z}$, $R^2$, ${\delta}_{0.3}$, $O$ and running $time$. The optimal model parameters and performance are listed in Table~3.  As described in Table~3, the best performance is obtained on the sample BSW for these three methods, $MSE$, $MAE$, $Bias$, ${\sigma}_{\rm NMAD}$, ${\sigma}_{{\Delta}z}$, $R^2$, ${\delta}_{0.3}$ and $O$ respectively amount to 0.1576, 0.2649, -0.0006, 0.0999, 0.3971, 0.6680, 93.78, 20.45 for CatBoost; 0.1617, 0.2669, -0.0009, 0.1001, 0.4022, 0.6595, 93.61, 20.58 for XGBoost; 0.1626, 0.2686, 0.001, 0.1007, 0.4032, 0.6578, 93.50, 20.83 for Random Forest, respectively. Table~3 also shows that for the samples BS\_W, the performance with optical and infrared information are better than that only with optical information for any method. Comparison of the performance of XGBoost, CatBoost and Random forest in Table~3, CatBoost shows its superiority.

Then, with optimal model parameters and 80:20 training-test split, we train again for the samples BSW and BS\_W. The scatter figure and $\Delta{\rm z(norm)}$ distribution of estimated photometric redshifts and spectroscopic redshifts are shown in Figure~4. Figure~4 further proves that CatBoost outperforms XGBoost and Random Forest.

          \begin{table*}
          	\small
          	\begin{center}
          		\caption[]{The performance of photometric redshift estimation with the best features and optimal model parameters \label{tab:confusion}}
          		\begin{tabular}{p{9mm}p{15mm}<{\centering}p{10mm}<{\centering}p{22mm}<{\centering}p{7mm}<{\centering}p{7mm}<{\centering}p{10mm}<{\centering}p{7mm}<{\centering}p{7mm}<{\centering}p{7mm}<{\centering}p{7mm}<{\centering}p{7mm}<{\centering}p{6mm}<{\centering}}
          			\hline      			
          			Sample & Input pattern   & Method  &Model parameter&MSE & MAE& $\mathrm{Bias}$ & ${\sigma}_{\rm NMAD}$& ${\sigma}_{{\Delta}\rm z}$&$R^2$&${\delta}_{0.3}$(\%)&$O$(\%)&Time(s)\\
          			\hline
          			BSW    & \text{Pattern I}  &XGBoost & max\_depth=11 &0.1617 &0.2669&-0.0009&0.1001 &0.4022&0.6595&93.61&20.58&110\\
          			& & & n\_estimators=1000 &&&&&&&&&\\      		
          			BSW    & \text{Pattern II}   &CatBoost & depth=12 &0.1576 &0.2649&-0.0006&0.0999 &0.3971&0.6680&93.78&20.45&421 \\
          			& && iterations=4000 &&&&&&&&&\\
          			BSW   &	\text{Pattern III}    &RF& max\_depth=15 &0.1626 &0.2686&0.0001&0.1007 &0.4032&0.6578&93.50&20.83&1336 \\
          			& & &n\_estimators=500 &&&&&&&&&\\
          			\hline
          			
          			BS\_W    &  \text{Pattern IV}  &XGBoost & max\_depth=10 &0.1865 &0.2969&0.0001&0.1106 &0.4318&0.6195&92.62&23.72&325\\
          			& && n\_estimators=1200 &&&&&&&&&\\      		
          			BS\_W      &    \text{Pattern V}   &CatBoost & depth=12&0.1848 &0.2960&0.0001&0.1105 &0.4299&0.6228&92.69&23.55&468 \\
          			& & & iterations=4000 &&&&&&&&&\\
          			BS\_W      &    \text{Pattern VI}    &RF& max\_depth=15 &0.1893&0.2990&0.0009&0.1111 &0.4351&0.6136&92.44&23.82&1431 \\
          			& & & n\_estimators=500 &&&&&&&&&\\
          			\hline
          			
          			BS\_W    &      \text{Pattern VII}  & XGBoost & max\_depth=10& 0.2930 &0.3967&-0.0001&0.1626 &0.5413&0.4066&87.43&37.17&110\\
          			& && n\_estimators=1000 &&&&&&&&&\\      		
          			
          			BS\_W    &     \text{Pattern VII}   &CatBoost & depth=12&0.2906 &0.3970&-0.0001&0.1637 &0.5393&0.4110&87.47&37.38&254 \\
          			& & & iterations=2000 &&&&&&&&&\\
          			
          			BS\_W    &      \text{Pattern VII}    &RF& max\_depth=15 &0.2944&0.3975&0.0003&0.1623 &0.5425&0.4038&87.36&37.07&1124 \\
          			& & & n\_estimators=500 &&&&&&&&&\\   		
          			
          			\hline
          		\end{tabular}
          	\end{center}
          \end{table*}

          \begin{figure*}
          	\centering
          	{
          		\includegraphics[height=5cm,width=7cm]{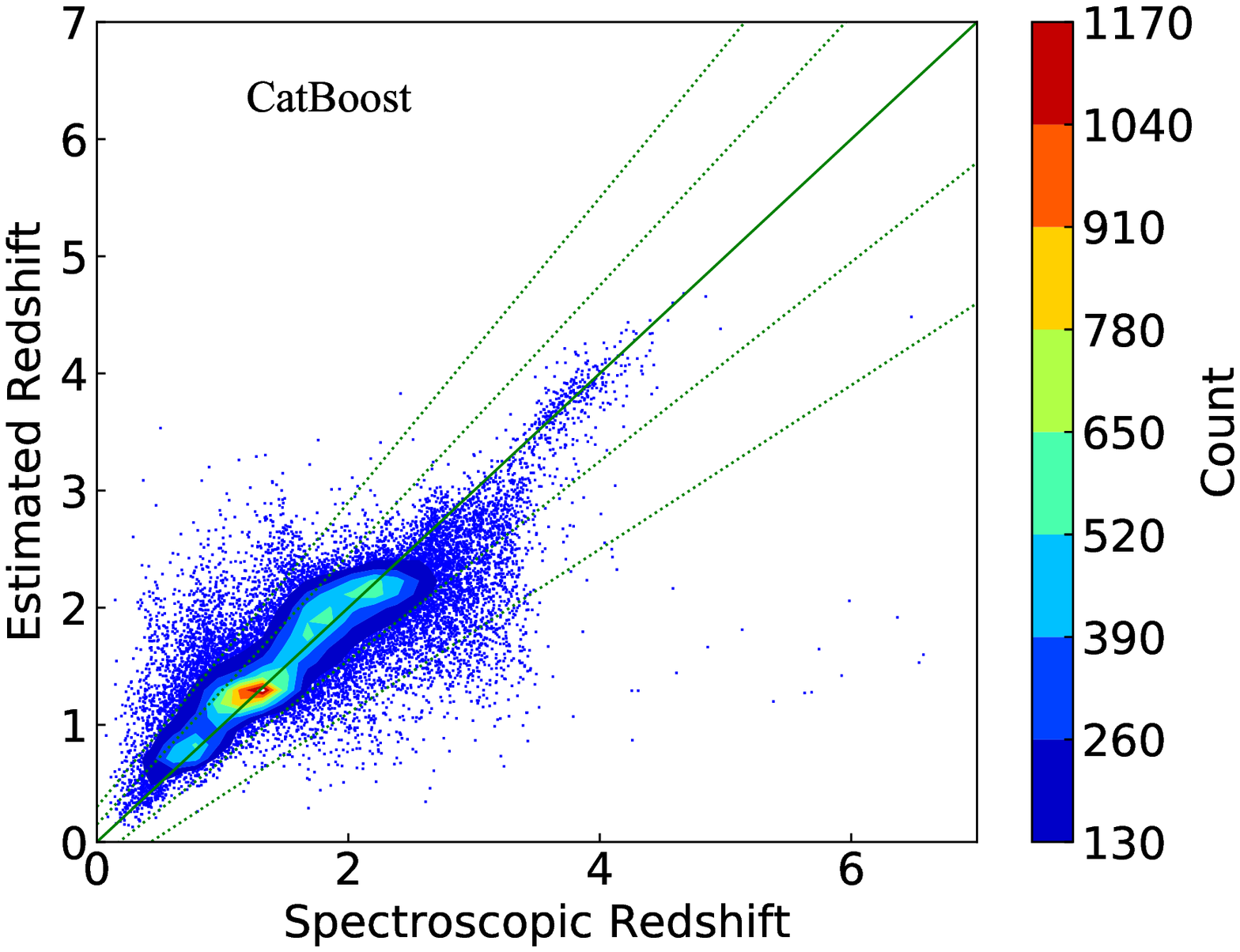}
          		\includegraphics[height=5cm,width=7cm]{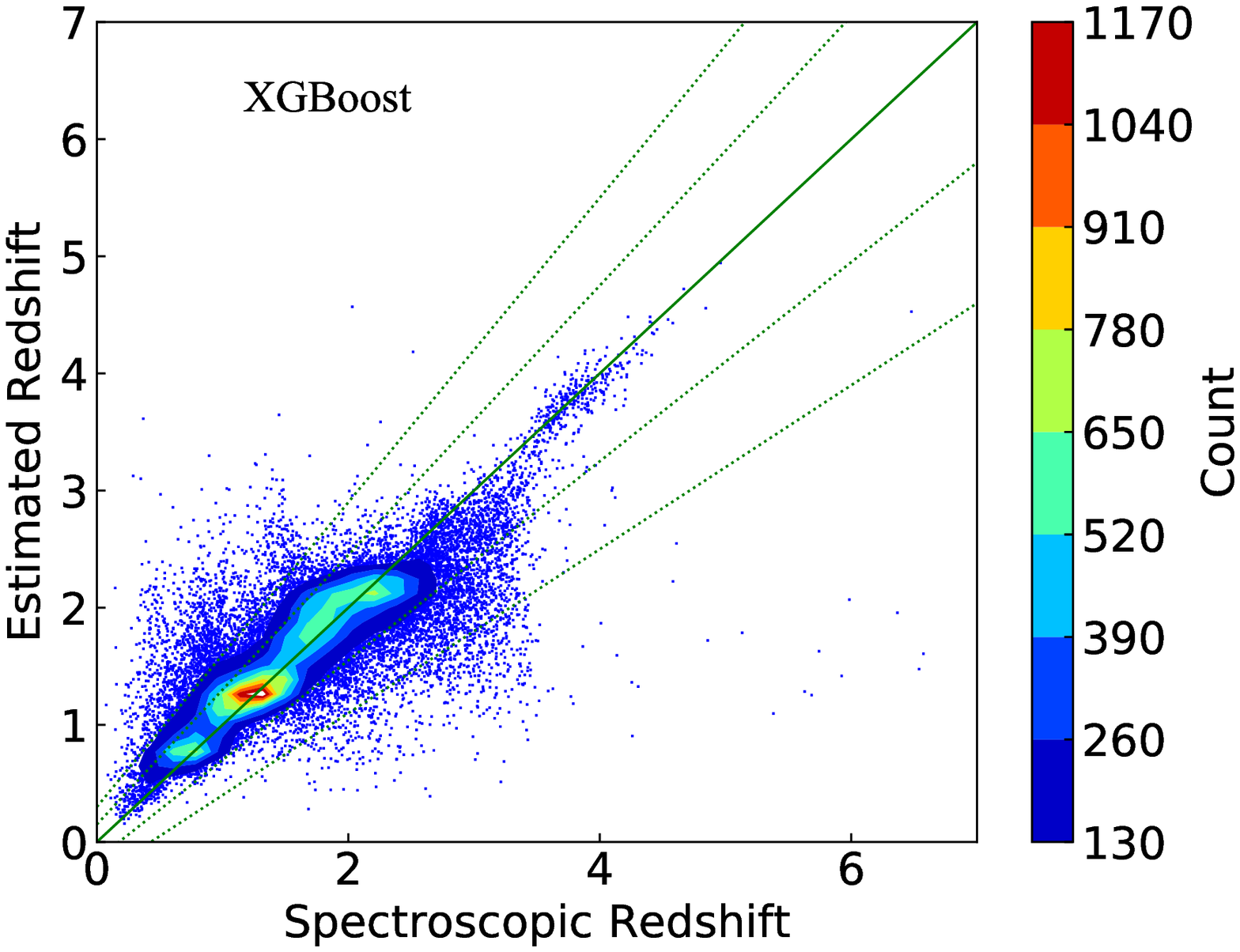}
          	}
          	\subfigure[For the sample BSW]{
          		\includegraphics[height=5cm,width=7cm]{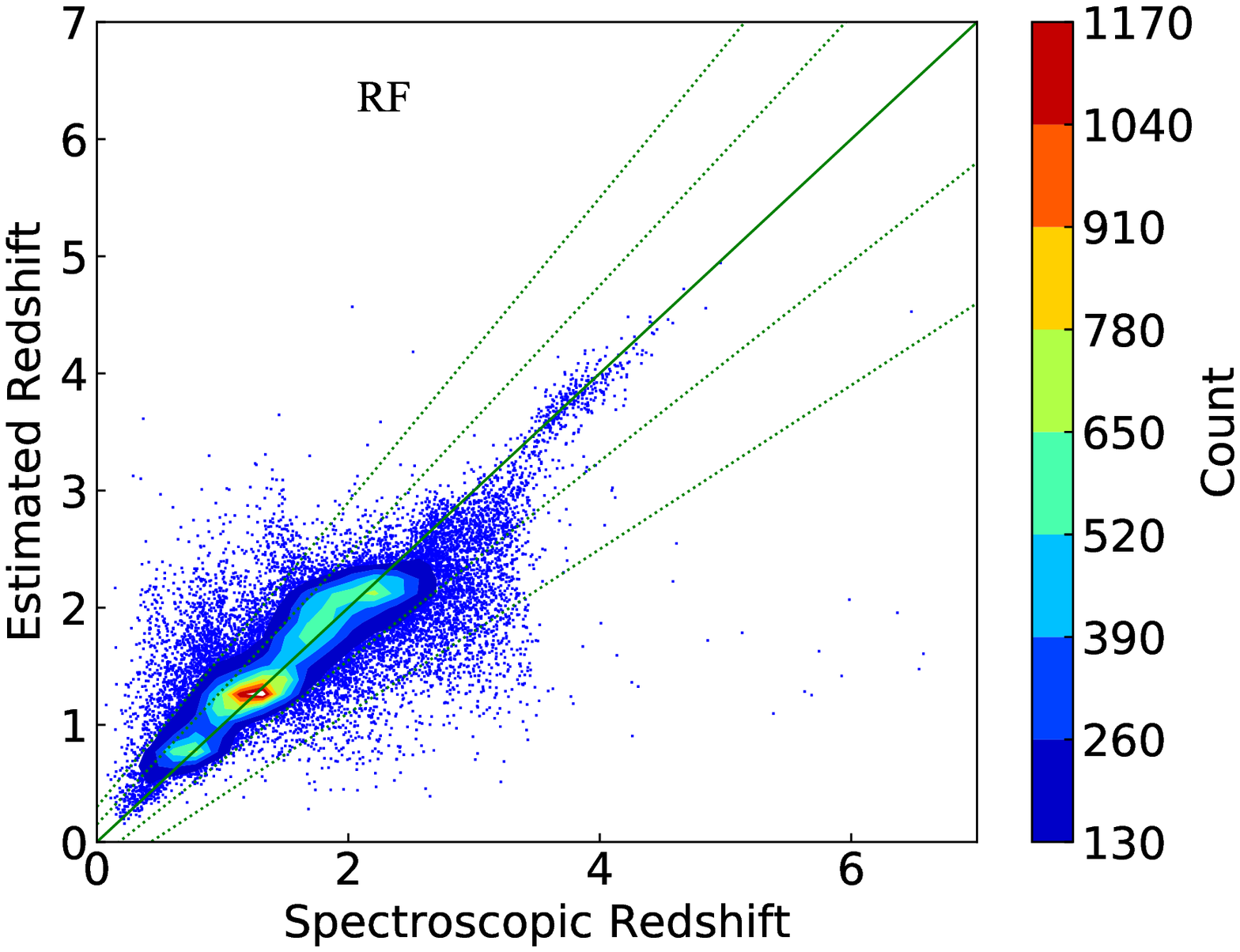}             		
          		\includegraphics[height=5cm,width=7cm]{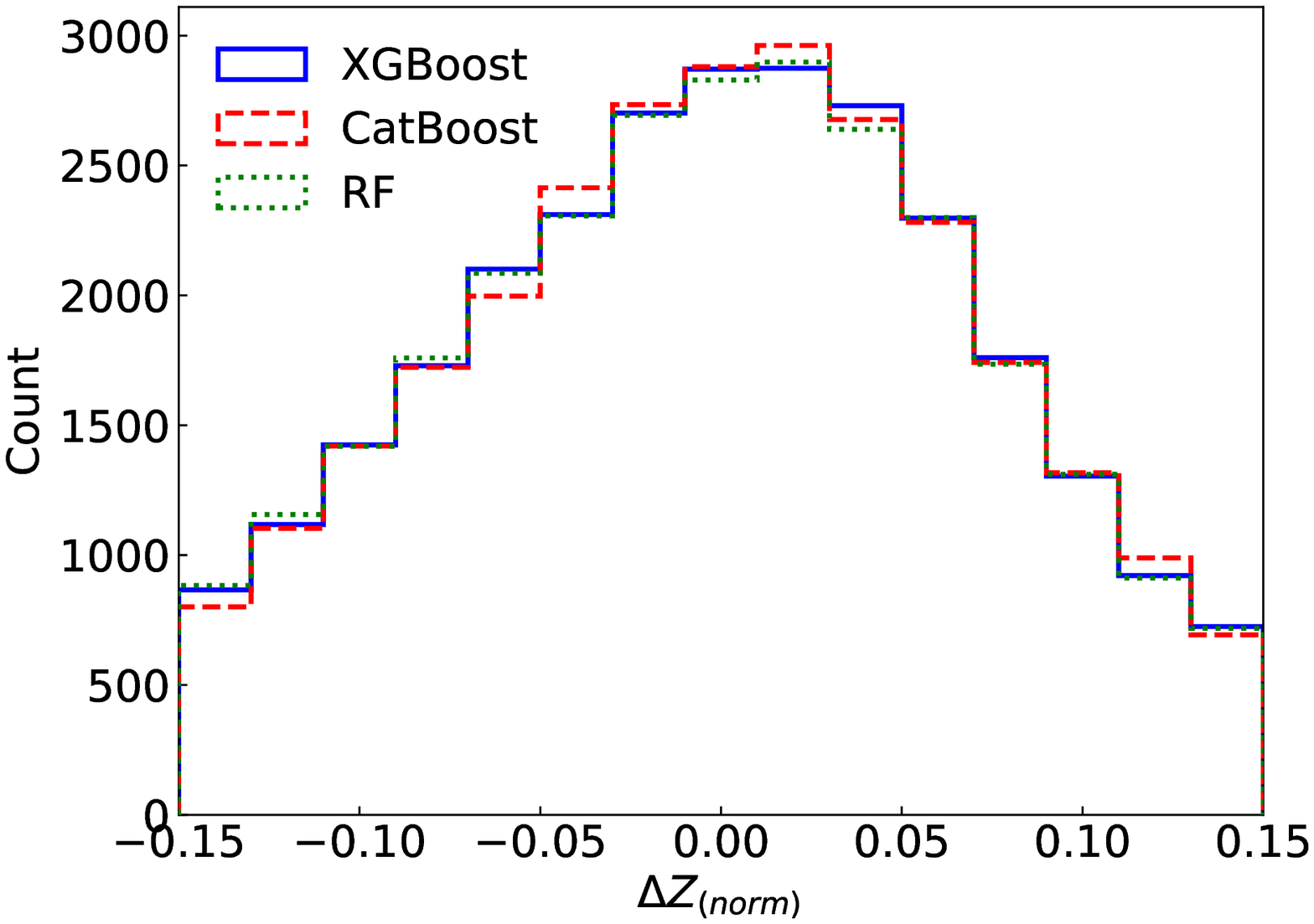}
          	}
          	{
          	\includegraphics[height=5cm,width=7cm]{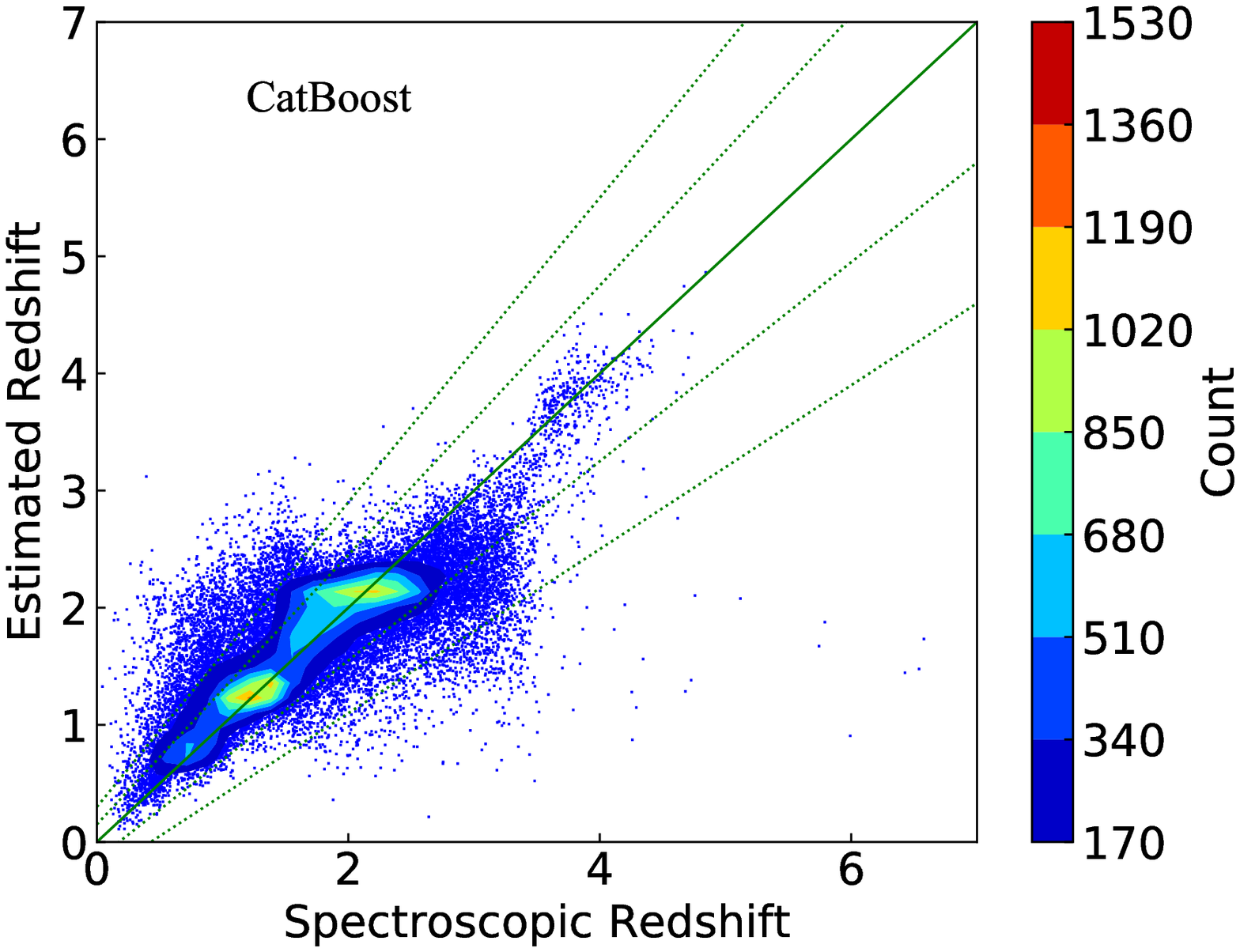}
          	\includegraphics[height=5cm,width=7cm]{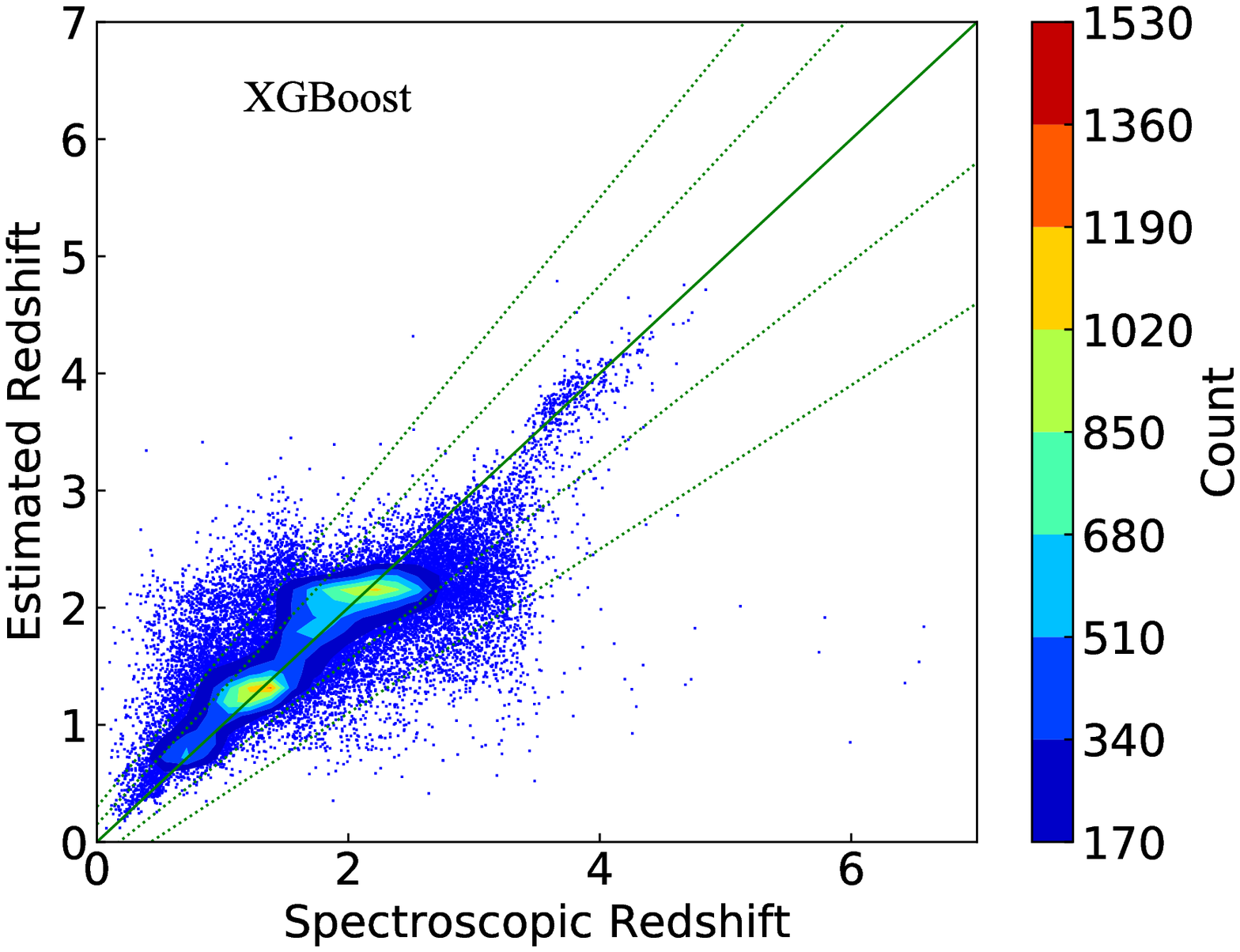} 	     	
          	}
          	\subfigure[For the sample BS\_W]{
          		\includegraphics[height=5cm,width=7cm]{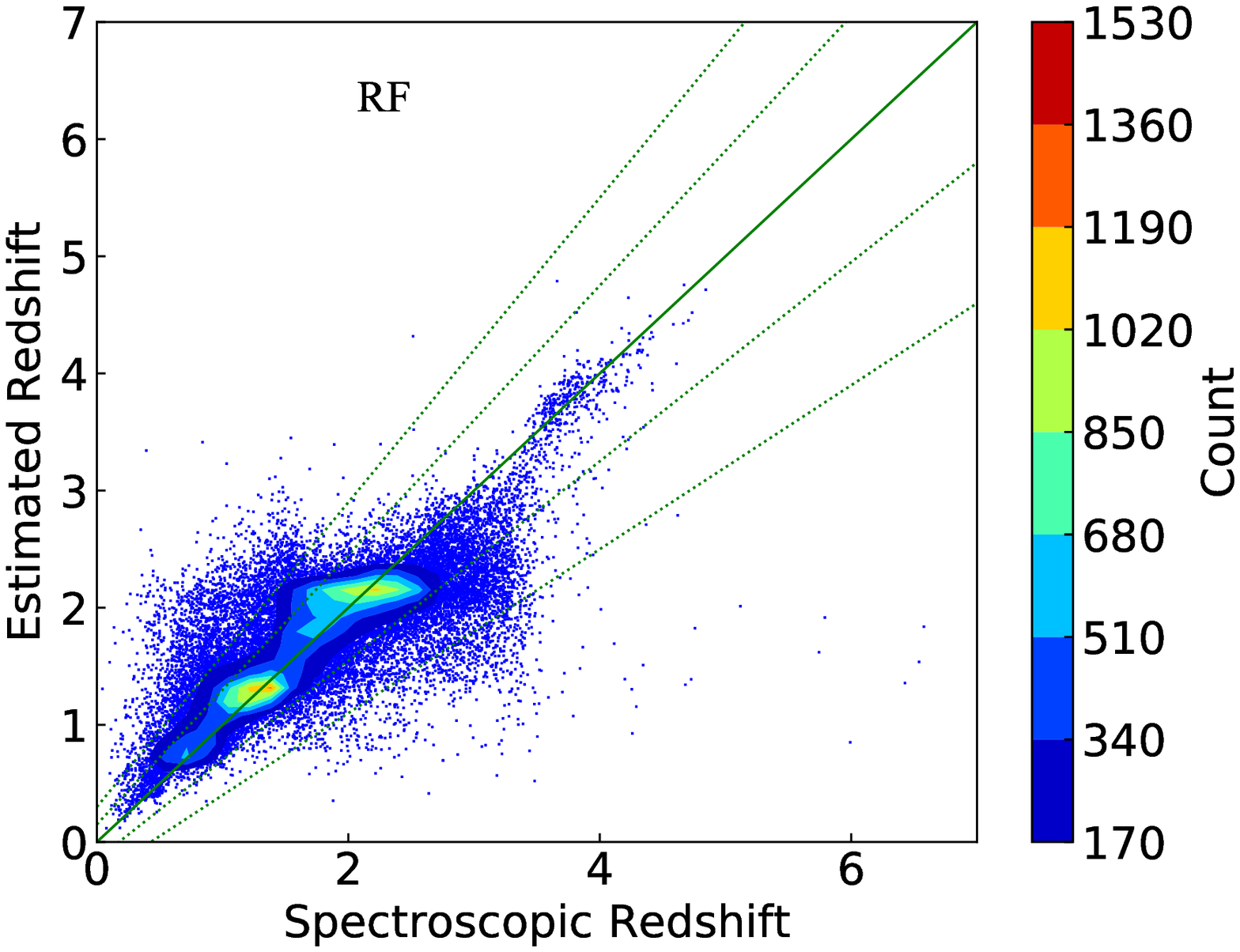}
          		\includegraphics[height=5cm,width=7cm]{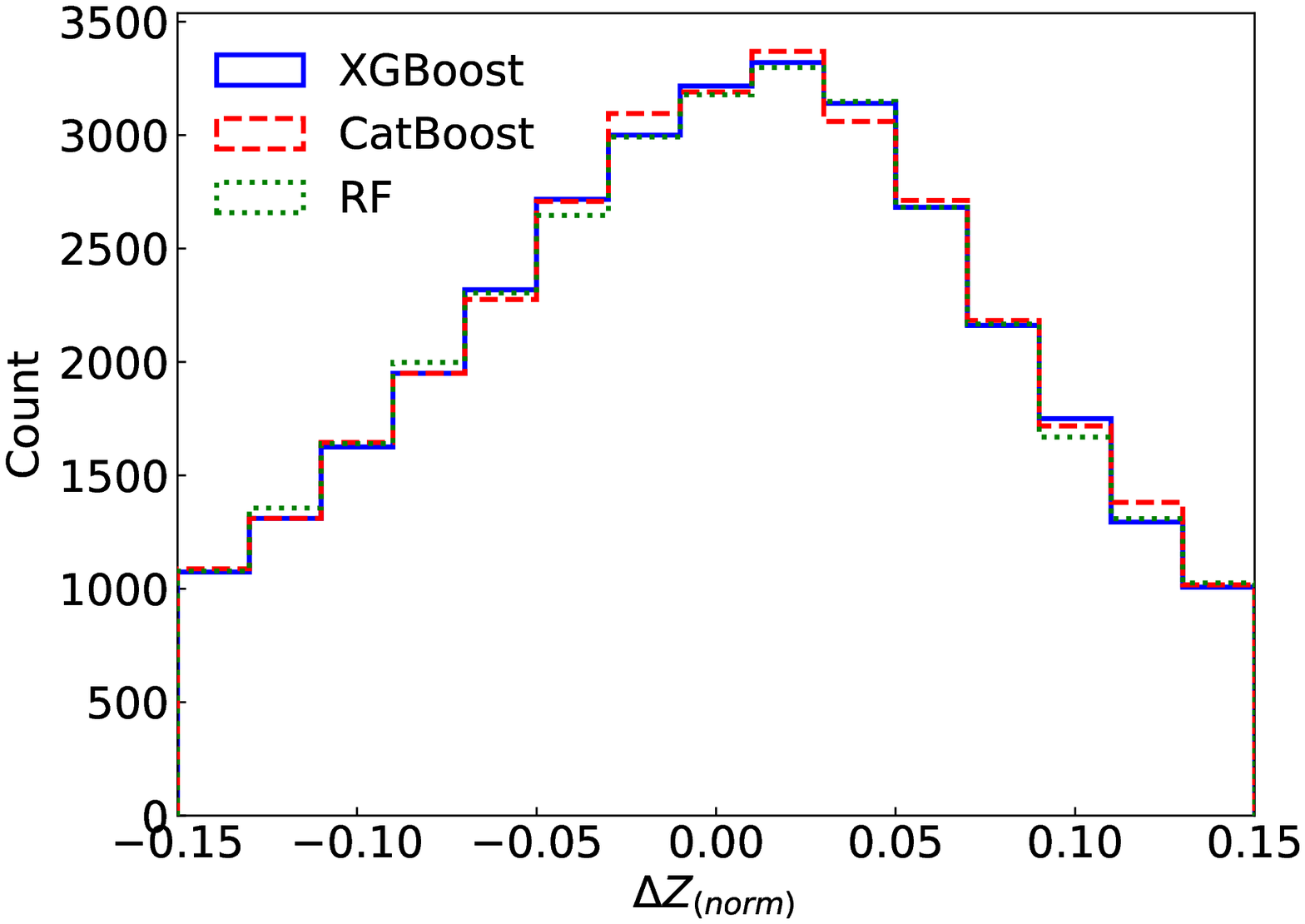}          	
          	}
          	\caption{The first three plots in (a) and (b) show the performance of photometric redshift estimation with CatBoost, XGBoost and RF, respectively. The green line represents $\Delta \rm z=0$; the green dotted lines
represent $\Delta \rm z(norm)=\pm 0.15, \pm 0.3$, separately. The last plots in (a) and (b) show the distribution of $\Delta \rm z(norm)$.}
          	\label{fig4}
          \end{figure*}

Finally, we use the three optimal models to train regressors with the total samples, and apply the total samples and external samples as test samples separately to validate the regressors, the validation performance is shown in Table~4. Table~4 tells that when the regressors created with optical and infrared information is applied, the performance of CatBoost with Pattern II is $MSE=0.1059$, $MAE=0.2223$, $Bias=-1.6\times 10^{-5}$, ${\sigma}_{\rm NMAD}=0.0872$, ${\sigma}_{{\Delta}z}=0.3254$, $R^2=0.7780$, ${\delta}_{0.3}=96.01\%$ and $O=15.79\%$ for the test sample BSW; the performance of CatBoost with Pattern II is with $MSE=0.1239$, $MAE=0.2134$, $Bias=0.0265$, ${\sigma}_{\rm NMAD}=0.0797$, ${\sigma}_{{\Delta}z}=0.3520$, $R^2=0.7585$, ${\delta}_{0.3}=94.50\%$ and $O=15.11\%$ for the test sample BLW. For the test sample BL\_W or BSO, the regressor with Pattern V is better than that with Pattern VII, which suggests that the regressor with Pattern V had better be adopted although the sources to be predicted only contain optical information. For the test sample BLW or BSW, the regressor with Pattern II is superior to that with Pattern V. Therefore the sources with optical and infrared information had better be estimated by the regressor with Pattern II. Table~4 further shows that the CatBoost regressors are effective to predict photometric redshifts of quasars.

        \begin{table*}
        	\begin{center}
        		\caption[]{The performance of CatBoost for photometric redshift estimation.  \label{tab:confusion}}
        		\begin{tabular}{rcccccccccc}
        			\hline
        			Training Sample & Input pattern  & Test Sample&MSE & MAE& $\mathrm{Bias}$ & ${\sigma}_{\rm NMAD}$& ${\sigma}_{{\Delta}\rm z}$&$R^2$&${\delta}_{0.3}$(\%)&$O$(\%)\\
        			\hline
        			BSW      & \text{Pattern  II}&BSW   &0.1059&0.2223&-1.6 $\times 10^{-5}$&0.0872&0.3254&0.7780&96.01&15.79\\
        			BSW      & \text{Pattern  II}&BLW   &0.1239&0.2134&0.0265&0.0797&0.3520&0.7585&94.50&15.11\\
        			BS\_W    & \text{Pattern  V} &BSW   &0.1115&0.2279&-4.0 $\times 10^{-5}$&0.0889&0.3340&0.7661&95.77&16.42\\
        			BS\_W    & \text{Pattern  V} &BS\_W &0.1365&0.2578&-7.0 $\times 10^{-6}$&0.0981&0.3695&0.7239&94.84&19.42\\
        			BS\_W    & \text{Pattern  V} &BLW   &0.1265&0.2167&0.0268&0.0808&0.3557&0.7534&94.48&15.61\\
        			BS\_W    & \text{Pattern  V} &BL\_W &0.1277&0.2188&0.0296&0.0817&0.3574&0.7502&94.41&15.87\\
        			BS\_W    & \text{Pattern  V} &BSO   &0.2490&0.3929&1.2 $\times 10^{-4}$&0.1550&0.4991&0.3884&90.64&32.93\\
        			BS\_W  & \text{Pattern  VII} &BSO   &0.3267&0.4363&-0.1853&0.1672&0.5719&0.1970&90.98&36.97\\
        			BS\_W  & \text{Pattern  VII} &BL\_W &0.2596&0.3681&0.1223&0.1639&0.5095&0.4925&85.82&37.61\\
        			\hline
        		\end{tabular}
        	\end{center}
        \end{table*}

\section{Two-step model}
In the whole quasar samples, the number of high redshift quasars (redshift $\ge 3.5$) is relatively small compared to that of low redshift quasars (redshift $< 3.5$). In the sample BS\_W, there are only 2238 high redshift quasars among the 213 359 quasars, while in the sample BSW, there are only 1798 high redshift quasars among 174 645 quasars. The sample BS\_W is divided into high redshift subsample BS\_W\_H and low redshift subsample BS\_W\_L, similarly the sample BSW is split into BSW\_H and BSW\_L. Table~5 shows the performance of photometric redshift estimation for high redshift and low redshift subsamples based on the best CatBoost regressors in Table~4. As described in Table~5, the performance on the high redshift subsamples is much worse than that on the low redshift subsamples for the sample BS\_W or BSW. Meanwhile, Figure~4 indicates that many high redshift quasars have been predicted as low redshift quasars. For the sample BS\_W, there are only 1765 quasars which are correctly predicted as high redshift quasars among 2238 high redshift quasars, taking up about 78 per cent, while for the sample BSW, 1492 quasars are correctly predicted as high redshift quasars among 1798, occupying about 83 per cent. It is obvious that the regressor fit for low redshift quasars is not fit for high redshift quasars. Thus, we propose a two-step model to improve the performance of redshift estimation for quasars, especially for high redshift quasars.

 \begin{table*}
 	\begin{center}
 		\caption[]{The performance of photometric redshift estimation for high and low subsamples with the best CatBoost regressors in Table~4. \label{tab:performance}}
 		\begin{tabular}{rcccccccccc}
 			\hline
 			Training Sample &	Pattern   &Test Sample & MSE & MAE& $\mathrm{Bias}$ & ${\sigma}_{\rm NMAD}$& ${\sigma}_{{\Delta}\rm z}$&$R^2$&${\delta}_{0.3}$(\%)&$O$(\%)\\
 			\hline
 			BS\_W &	\text{Pattern  V}   & BS\_W\_H  & 0.4079& 0.2799 &-0.1943&0.0359&0.6387&-1.4640&96.15&8.04 \\
 			BS\_W &	\text{Pattern  V}   & BS\_W\_L  & 0.1336&0.2575  &0.0021&0.0989&0.3655&0.7019&94.83&19.34 \\
 			\hline
 			BSW & 	\text{Pattern  II}   &BSW\_H   & 0.4074& 0.2575&-0.1741&0.0318&0.6383&-1.1900&96.27&6.7 \\
 			BSW & 	\text{Pattern  II}   &BSW\_L   & 0.1027 &0.2219&0.0018&0.0879&0.3205&0.7591&96.00&15.89 \\
 			\hline
 		\end{tabular}
 	\end{center}
 \end{table*}

\subsection{The first step of the two-step model}
In order to improve the performance of photometric redshift estimation of quasars, we put forward a new scheme of first classification and second regression (i.e. two-step model) for photometric redshift estimation. The first step of two-step model is to construct a classifier to discriminate the whole quasars into low and high redshift subsamples. Based on the samples BSW and BS\_W, we add a column ``label", which is 1 if redshift $\ge$ 3.5, otherwise, set to 0. We adopt four standard metrics (Accuracy, Precision, Recall, and F1\_score) to evaluate the performance of one classifier.
Accuracy (Accu.) is the ratio of the number of the correctly classified samples to the total number of the samples.
\begin{equation}
\mathrm{Accuracy} = \frac{TP+TN}{TP + TN + FP + FN}
\end{equation}
Here TP and TN are correctly classified by the classifier. FP shows the negative sample classified as positive. FN represents the positive sample classified as negative.

Precision (Prec.) is the ratio of the true positive (negative) sample in all the samples that are classified as positive (negative).
Recall (Rec.) is the ratio of correctly classified positive (negative) samples in all the true positive (negative) samples.
\begin{equation}
\mathrm{Precison} = \frac{TP}{TP + FP}, \mathrm{Recall} = \frac{TP}{TP + FN}
\end{equation}

F1\_score (F1) is a weighted average of Precision and Recall.
\begin{equation}
\mathrm{F1\_score} = \frac{2 \times (\mathrm{Prec}. \times \mathrm{Rec.})}{\mathrm{Prec.} + \mathrm{Rec.}}
\end{equation}

Then we use CatBoost, XGBoost and Random Forest to construct binary classifiers, the optimal performance and hyper parameters of the three classifiers by five-fold validation are shown in Table~6. Table~6 shows that CatBoost is superior to XGBoost and Random Forest when separating the sample BSW or BS\_W into low and high redshift subsamples considering Accuracy and F1\_score. Moreover Recall (completeness) of CatBoost achieves the best for high redshift quasars. In order to find more high redshift quasars, we should keep high completeness at high redshift in the first step of two-step model.

             \begin{table*}
             	\begin{center}
             		\caption[]{The performance of different classifiers for high and low redshift subsamples. }
             		\begin{tabular}{p{8mm}<{\centering}p{18mm}<{\centering}p{12mm}<{\centering}p{22mm}<{\centering}p{10mm}<{\centering}p{9mm}<{\centering}p{9mm}<{\centering}p{8mm}<{\centering}p{9mm}<{\centering}p{9mm}<{\centering}p{8mm}<{\centering}p{7mm}<{\centering}}       		
             			\hline
             			&      &       & &&\multicolumn{3}{c}{High redshift}&\multicolumn{3}{c}{Low redshift}&\\
             			\hline
             			Sample&Input Pattern&Method&Parameter&Accu.(\%)&Prec.(\%)&Rec.(\%)&F1(\%)&Prec.(\%)&Rec.(\%)&F1(\%)&Time(s)\\
             			\hline
             			BSW  &	Pattern I   &XGBoost &max\_depth=6  &99.78  &92.48  &86.10 &89.18 &99.86  &99.92 &99.89 &124 \\
             			&&&n\_estimators=200&&&&&&&& \\
             			BSW  &	Pattern II    &CatBoost &depth=6  &\textbf{99.79}  &92.33  &\textbf{87.04} &\textbf{89.60}  &\textbf{99.87}  &99.93  &\textbf{99.90} &19 \\
             			&&&iterations=1000&&&&&&&& \\
             			BSW  &	Pattern III   &RF  &max\_depth=14 &99.78 &\textbf{93.91}  &84.54 &88.98 &99.84  &\textbf{99.94}  &99.89&190 \\
             			&&&n\_estimators=500&&&&&&&& \\
             			\hline
             			BS\_W  &	Pattern IV   &XGBoost &max\_depth=6  &99.77  &92.30  &85.12 &88.56 &\textbf{99.84}  &99.92 &99.87 & 46 \\
             			&&&n\_estimators=100&&&&&&&& \\
             			BS\_W  &	Pattern V  &CatBoost  &depth=12  &\textbf{99.78}  &92.87  &\textbf{85.21} &\textbf{88.88} &\textbf{99.84}  &99.93 &\textbf{99.88}  & 130 \\
             			&&&iterations=1000&&&&&&&& \\
             			BS\_W  &	Pattern VI &RF  & max\_depth=13&99.77 &\textbf{93.88}  &82.89 &88.04 &99.82  &\textbf{99.94} &99.87 &72 \\
             			&&&n\_estimators=100&&&&&&&& \\
             			\hline              				            				          			
             		\end{tabular}
             	\end{center}
             \end{table*}

In order to further test the performance of CatBoost, XGBoost and Random Forest classifiers, we use the whole quasar samples from SDSS as training and test samples. The self-validation classification performance of the three classifiers is shown in Table~7. It can be seen from Table~7 that the CatBoost classifier based on the sample BS\_W has better performance than that based on the sample BSW when the sample BSW is taken as test sample. The better performance is that Accuracy reaches 99.99 per cent, F1 is 99.96 per cent for the high redshift subsample, and F1 is 99.99 per cent for the low redshift subsample. Therefore the CatBoost classifier based on the sample BS\_W is more suitable for the sample BSW. In other words, the CatBoost classifier created on the training sample BS\_W had better be adopted for the sources with optical information or those with optical and infrared information. When the BSW sample is taken as training and test samples, XGBoost achieves the best performance while CatBoost and Random Forest obtain the same performance (all metrics above 96.27 per cent). In other two situations, CatBoost is superior to XGBoost and Random Forest, moreover the running time of CatBoost is the fastest. Therefore CatBoost is adopted as the core algorithm of classification in the first step of the whole scheme.

     \begin{table*}
     	\begin{center}
     		\caption[]{The performance of different classifiers with different training and test samples. }
     		\begin{tabular}{p{12mm}cp{15mm}p{12mm}cp{9mm}p{9mm}p{9mm}p{9mm}p{9mm}p{9mm}c}
     			\hline
     			&       & &&&\multicolumn{3}{c}{High redshift}&\multicolumn{3}{c}{Low redshift}&\\
     			\hline
     			Training Sample&Method& Input \quad pattern&Test Sample&Accu.(\%)&Prec.(\%)&Rec.(\%)&F1(\%)&Prec.(\%)&Rec.(\%)&F1(\%)&Time(s)\\
     			\hline               			
     			BSW    &CatBoost&	Pattern II   &BSW    &99.96  &100  &96.27 &98.10  &99.96  &100 &99.98  &0.25 \\
     			BS\_W  &CatBoost&	Pattern V    &BSW    &99.99  &100  &99.94 &99.97  &99.99  &100  &100 &0.27 \\
     			BS\_W  &CatBoost&	Pattern V    &BS\_W  &99.99  &100  &99.91 &99.95 &99.99  &100  &100 &0.36 \\
     			\hline               			
     			BSW    &XGBoost&	Pattern I   &BSW    &100  &100  &100 &100  &100  &100 &100  &2 \\
     			BS\_W  &XGBoost&	Pattern IV    &BSW    &99.97  &99.60  &98.00 &98.79  &99.98  &99.99  &99.99 &1.5 \\
     			BS\_W  &XGBoost&	Pattern IV    &BS\_W  &99.97  &99.68  &97.90 &98.78 &99.98  &99.99  &99.99 &2 \\                 			
     			\hline              				            				          			
     			BSW    &RF&	Pattern III   &BSW    &99.96  &100  &96.27 &98.10  &99.96  &100 &99.98  &11 \\
     			BS\_W &RF &	Pattern VI    &BSW    &99.93  &99.53  &94.16 &96.77  &99.94  &99.99  &99.96 &3 \\
     			BS\_W &RF &	Pattern VI    &BS\_W  &99.92  &99.33  &93.07 &96.10 &99.93  &99.99  &99.96 &3 \\
     			\hline
     		\end{tabular}
     	\end{center}
     \end{table*}

\subsection{The second step of the two-step model}
High and low redshift subsamples of BSW and BS\_W are trained to construct new regressors by CatBoost, XGBoost and Random Forest, respectively. The optimal model parameters and performance of all subsamples for these three methods are shown in Table~8. As shown in Table~8, the performance of both low and high redshift subsamples has been significantly improved, especially for high redshift quasars. Taking the evaluation metrics of regression into account except Bias, ${\sigma}_\mathrm{NMAD}$ and $O$ in Table~8, CatBoost outperforms XGBoost and Random Forest for any sample. Only given ${\sigma}_\mathrm{NMAD}$, CatBoost obtains the best performance for the samples BSW\_L and BS\_W\_L. Only considering $O$, CatBoost shows its superiority for the samples BSW\_H, BSW\_L and BS\_W\_L. As a result, CatBoost is taken as the core regression algorithm for the two-step model.

\begin{table*}
	\begin{center}
		\caption[]{The performance of photometric redshift estimation on different subsamples with different methods.\label{tab:performance}}
		\begin{tabular}{p{9mm}p{15mm}<{\centering}p{10mm}<{\centering}p{22mm}<{\centering}p{7mm}<{\centering}p{7mm}<{\centering}p{10mm}<{\centering}p{7mm}<{\centering}p{7mm}<{\centering}p{7mm}<{\centering}p{7mm}<{\centering}p{7mm}<{\centering}p{6mm}<{\centering}}
			\hline
			Sample & Input pattern   & Method  &Model parameter&MSE & MAE& $\mathrm{Bias}$ & ${\sigma}_\mathrm{NMAD}$& ${\sigma}_\mathrm{{\Delta}\rm z}$&$R^2$&${\delta}_{0.3}$(\%)&$O$(\%)&Time(s)\\
			\hline
			BSW\_H  &	\text{Pattern  I} &XGBoost &max\_depth=6 &0.0989  &0.1515&-0.0254&\textbf{0.0253}&0.3144&0.4777&99.11&2.56&2\\
			&&&n\_estimators=500&&&&&&&&&               				\\
			BSW\_H  &	\text{Pattern  II} &CatBoost  & depth=7  & \textbf{0.0756} &\textbf{0.1439}&-0.0055&0.0256&\textbf{0.2750}&\textbf{0.5910}&\textbf{99.72} &\textbf{2.17} &5\\
			&&&iterations=3000&&&&&&&&& \\
			BSW\_H  &	\text{Pattern  III} &RF  & max\_depth=11 &0.0781 & 0.1442&0.0035&\textbf{0.0253}&0.2796&0.5804&99.55 &2.4 & 14\\
			&&&n\_estimators=1000&&&&&&&&& \\
			\hline
			
			BSW\_L  &		\text{Pattern  I}&XGBoost  &max\_depth=10 &0.1526 &0.2646&-0.0017&0.1012&0.3907 &0.6408&93.77&20.64&32 \\
			&&&n\_estimators=700&&&&&&&&& \\			
			
			BSW\_L  &		\text{Pattern  II} &CatBoost  &depth=13  &\textbf{0.1497} & \textbf{0.2624}&-0.0005&\textbf{0.1002}&\textbf{0.3870}&\textbf{0.6475}&\textbf{93.85}&\textbf{20.37} & 285\\
			&&&iterations=3000&&&&&&&&& \\			
			
			BSW\_L  &		\text{Pattern  III} &RF   &max\_depth=15& 0.1536 & 0.2656&0.0010&0.1012&0.3919&0.6386&93.67 &20.68&1300 \\
			&&&n\_estimators=500&&&&&&&&& \\
			\hline
			BS\_W\_H &	\text{Pattern  IV}  &XGBoost &max\_depth=5   &0.0800&0.1429&-0.0068&0.0250&0.2828&0.5119&99.51&\textbf{2.24}&3\\
			&&&n\_estimators=1000&&&&&&&&&               				\\
			
			BS\_W\_H &	\text{Pattern  V}   &CatBoost&depth=6   & \textbf{0.0759}&\textbf{0.1404}&-0.0081&0.0248&\textbf{0.2756}&\textbf{0.5407}&\textbf{99.64}&2.32 &5\\
			&&&iterations=2000&&&&&&&&& \\
			
			BS\_W\_H &	\text{Pattern  VI}   &RF  &max\_depth=11 &0.0769& 0.1405&-0.0007&\textbf{0.0242}&0.2773&0.5285&99.51&2.59 & 5\\
			&&&n\_estimators=300&&&&&&&&& \\
			\hline
			BS\_W\_L &	\text{Pattern  IV}  &XGBoost &max\_depth=10  &0.1791&0.2947&-0.0002&0.1109 &0.4233&0.5967&92.72&23.66&370 \\
			&&&n\_estimators=1000&&&&&&&&& \\
			
			BS\_W\_L &	\text{Pattern  V}   &CatBoost &depth=13 &\textbf{0.1775}& \textbf{0.2934}&-0.0001&\textbf{0.1105} &\textbf{0.4213}&\textbf{0.6005}&\textbf{92.80}&\textbf{23.52} & 450\\
			&&&iterations=3000&&&&&&&&& \\
			
			BS\_W\_L &	\text{Pattern  VI}    &RF &max\_depth=15 &0.1812 & 0.2963&0.0006&0.1115&0.4257&0.5919&92.60 &23.76&2811 \\
			&&&n\_estimators=1000&&&&&&&&& \\
			\hline
		\end{tabular}
	\end{center}
\end{table*}

\subsection{Comparison of two-step model with one-step model}
For the two-step model, its first step is to classify the sample into low and high redshift subsamples; its second step is to create CatBoost regressors for low and high redshift subsamples respectively. According to the above experiments, CatBoost is regarded as the core algorithms for both classification and regression. The CatBoost classifier is trained on the BS\_W sample no matter whether the sources own infrared information, while the CatBoost regressors for low and high redshift subsamples are trained on the samples BS\_W\_L and BS\_W\_H for sources only with optical information and trained on the samples BSW\_L and BSW\_H with optical and infrared information. We use the samples BSW, BS\_W, BLW and BL\_W to test models. Table~4 gives the performance of photometric redshift estimation by one-step model. In order to compare the performance of photometric redshift estimation between one-step model and two-step model, the test results of the two models with different test samples are indicated in Table~9. Taking the samples BSW and BS\_W for example, the comparison of predicted photometric redshifts with spectroscopic redshifts is described in Figures~5 and the ${\Delta}{\mathrm{z}}$ distribution is indicated in Figure~6. In Figure~5, the left ones are the results of two-step model, the right ones are those of one-step model. As shown in Figure~5, the number of outliers for two-step model is less than that for one-step model, especially in the range of high redshift. From the evaluation metrics of regression in Table~9 and outliers in Figure~5, it is obvious that the performance of two-step model is better than that of one-step model. Figure~6 further proves this result.

                       \begin{table*}
                       	\begin{center}
                       		\caption[]{Comparison of the performance of photometric redshift estimation by two-step model with that by one-step model. \label{tab:confusion}}
                       		\begin{tabular}{rccccccccc}
                       			\hline
                       		     Test Sample &MSE & MAE & $\mathrm{Bias}$ & ${\sigma}_\mathrm{NMAD}$& ${\sigma}_\mathrm{{\Delta}z}$&$R^2$&${\delta}_{0.3}$(\%)&$O$(\%)&Time(s)\\
                       			\hline
                        &&&two-step model&&&&&&\\
                                \hline                     			
	                       		BSW    &0.0970 &0.2153&-0.00004&0.0854&0.3114&0.7967&96.32&15.13&25.0 \\
	                       		BLW    &0.1216 &0.2103&0.0249&0.0784&0.3487&0.7630&94.86&14.81&2.0 \\
	                       		BS\_W  &0.1266 &0.2499&-0.00004&0.0955&0.3558&0.7440&96.32&18.67&35.0 \\
	                       		BL\_W  &0.1251 &0.2157&0.0280&0.0804&0.3537&0.7553&94.40&15.63&9.0 \\
                                \hline
                         &&&one-step model&&&\\
                                \hline
                                BSW   &0.1059&0.2223&-0.00002&0.0872&0.3254&0.7780&96.01&15.80&0.80\\
                                BLW   &0.1239&0.2134&0.0265&0.0797&0.3521&0.7585&94.50&15.11&0.03\\
                                BS\_W &0.1365&0.2578&-0.000007&0.0981&0.3695&0.7239&94.84&19.42&1.00\\
                                BL\_W &0.1277&0.2188&0.0296&0.0817&0.3574&0.7502&94.40&15.88&0.05\\
                       			\hline
                       		\end{tabular}
                       	\end{center}
                       \end{table*}

                            \begin{figure*}
                            	\centering
                            	\subfigure[For the sample BSW. ]{
	                            	\includegraphics[height=6cm,width=7cm]{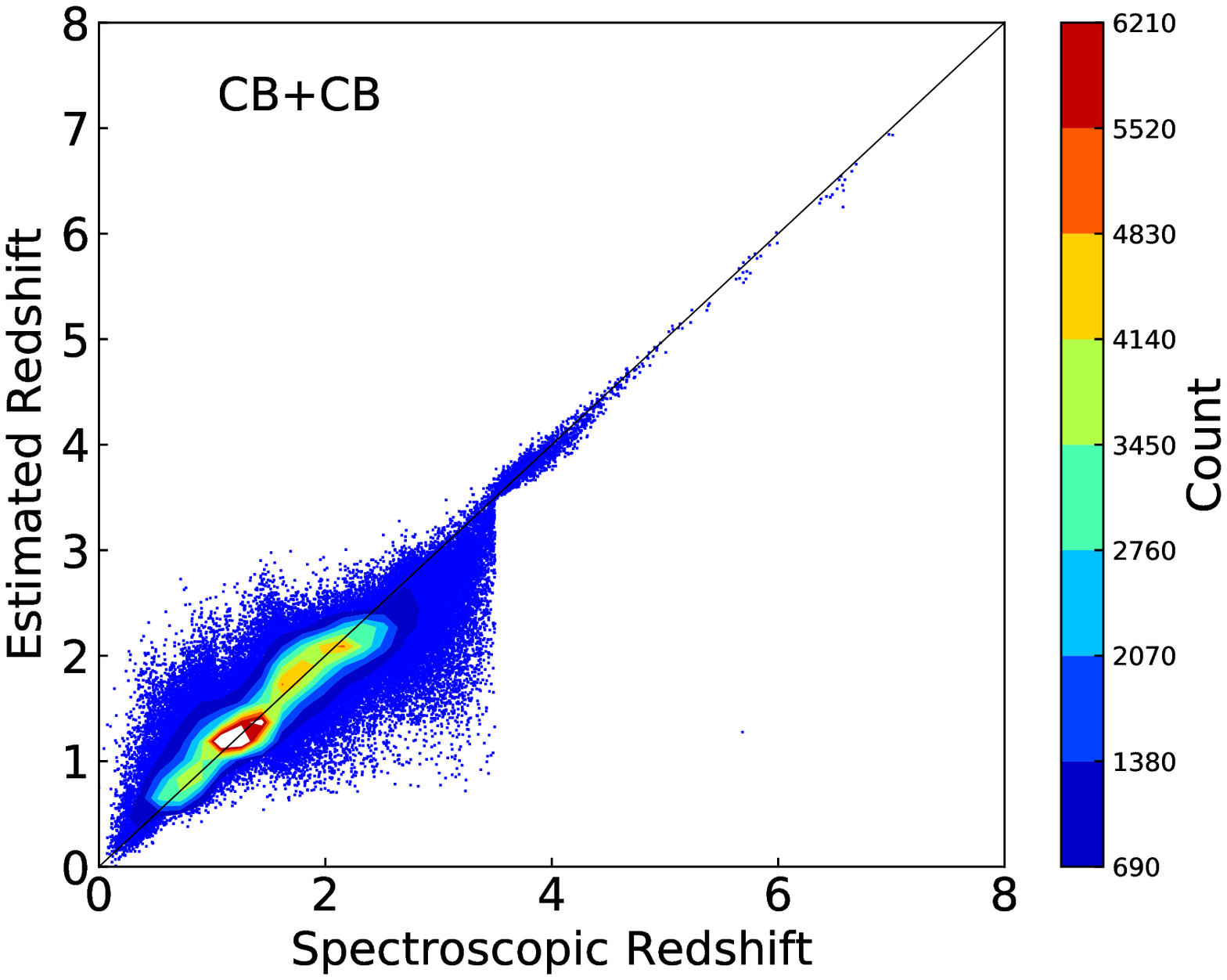}
	                            	\includegraphics[height=6cm,width=7cm]{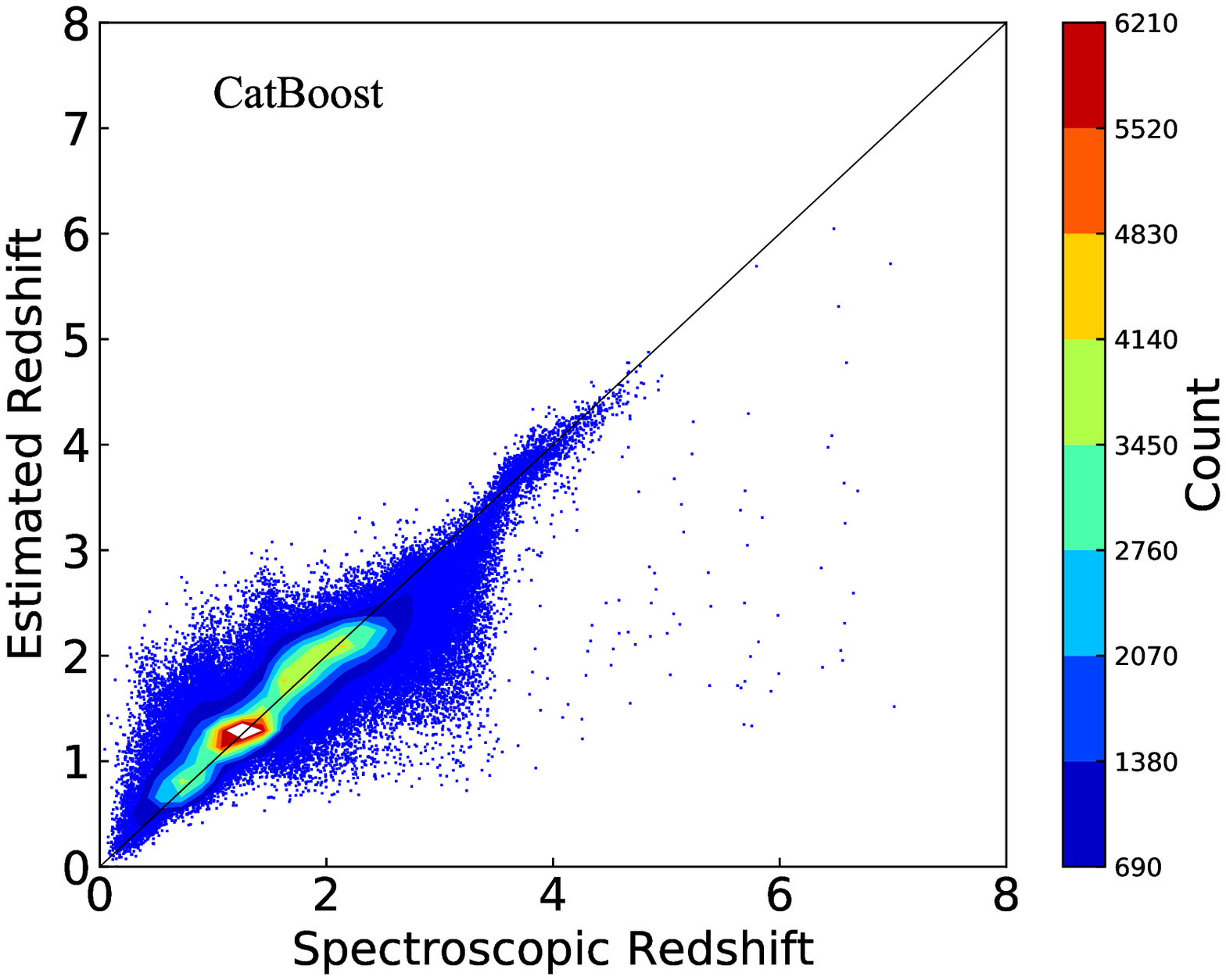}
	                            }
	                            \subfigure[For the sample BS\_W. ]{
	                            	\includegraphics[height=6cm,width=7cm]{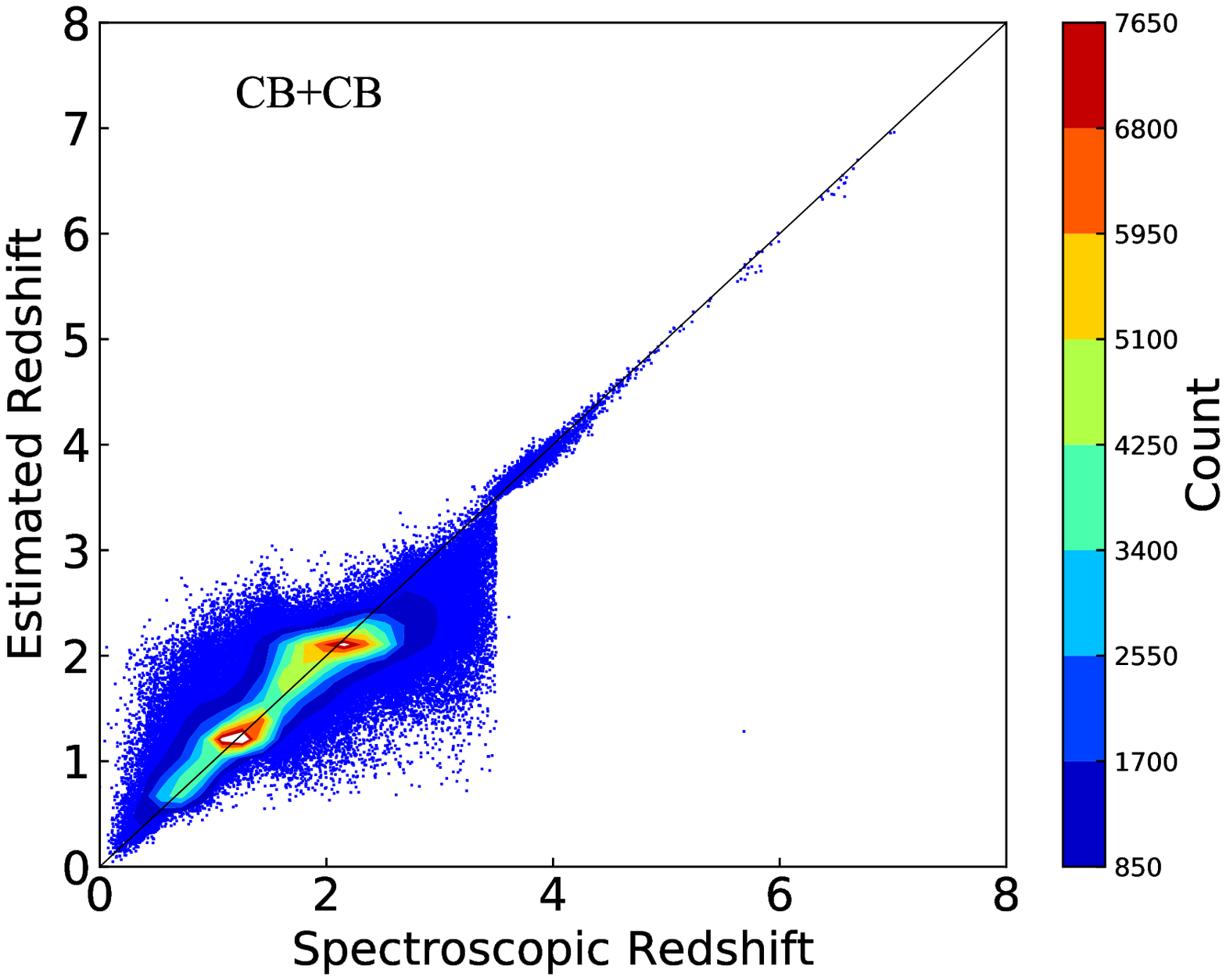}
	                            	\includegraphics[height=6cm,width=7cm]{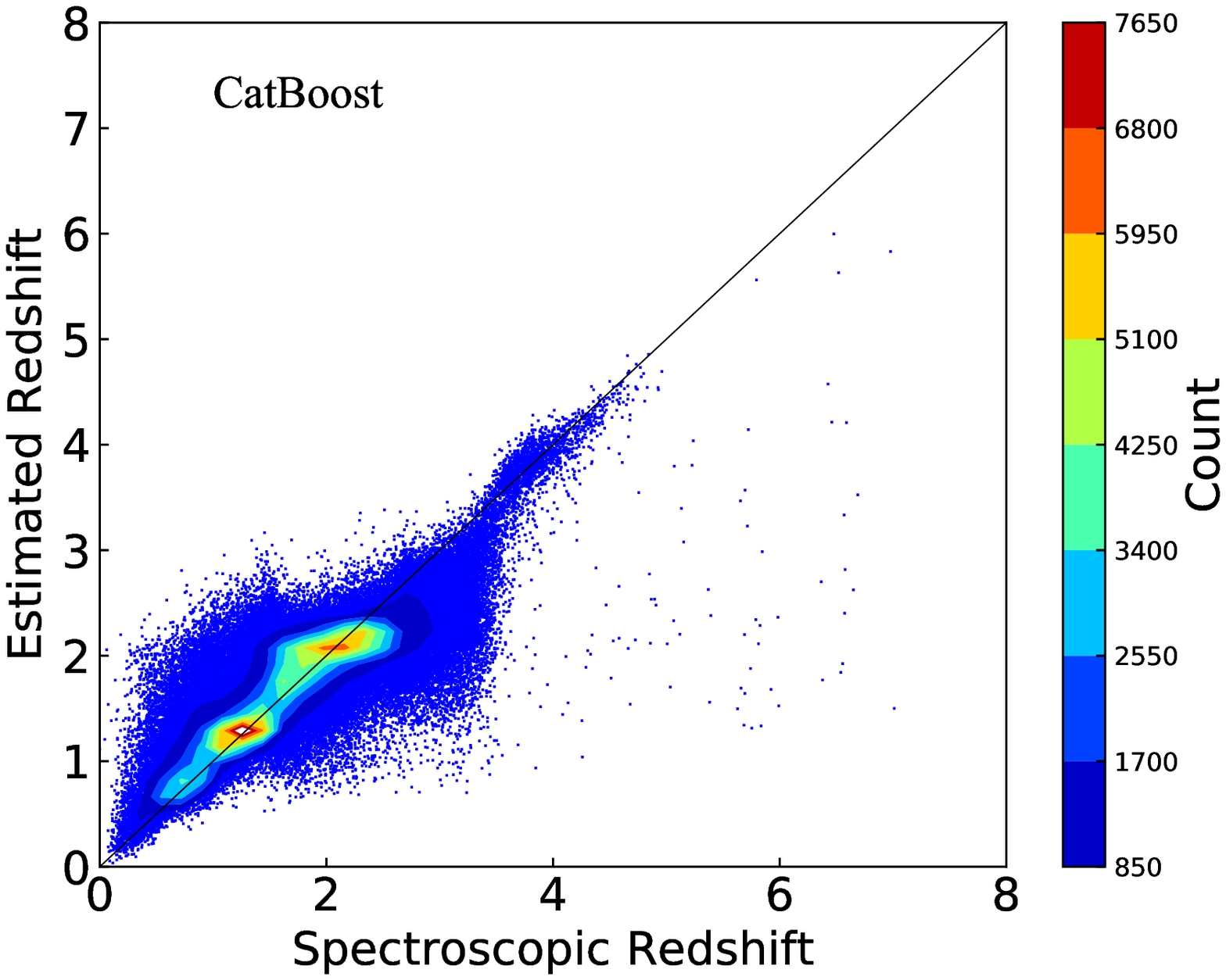}
	                            }
	
                            	\caption{Comparison of the photometric redshift with the spectroscopic redshift for the samples BSW and BS\_W with two-step models (left panel) and one-step models (right panel), respectively.}
                            	\label{fig5}
                            \end{figure*}

                                  \begin{figure*}
                                  	\centering
                                  	{
                                  		\includegraphics[height=6cm,width=7cm]{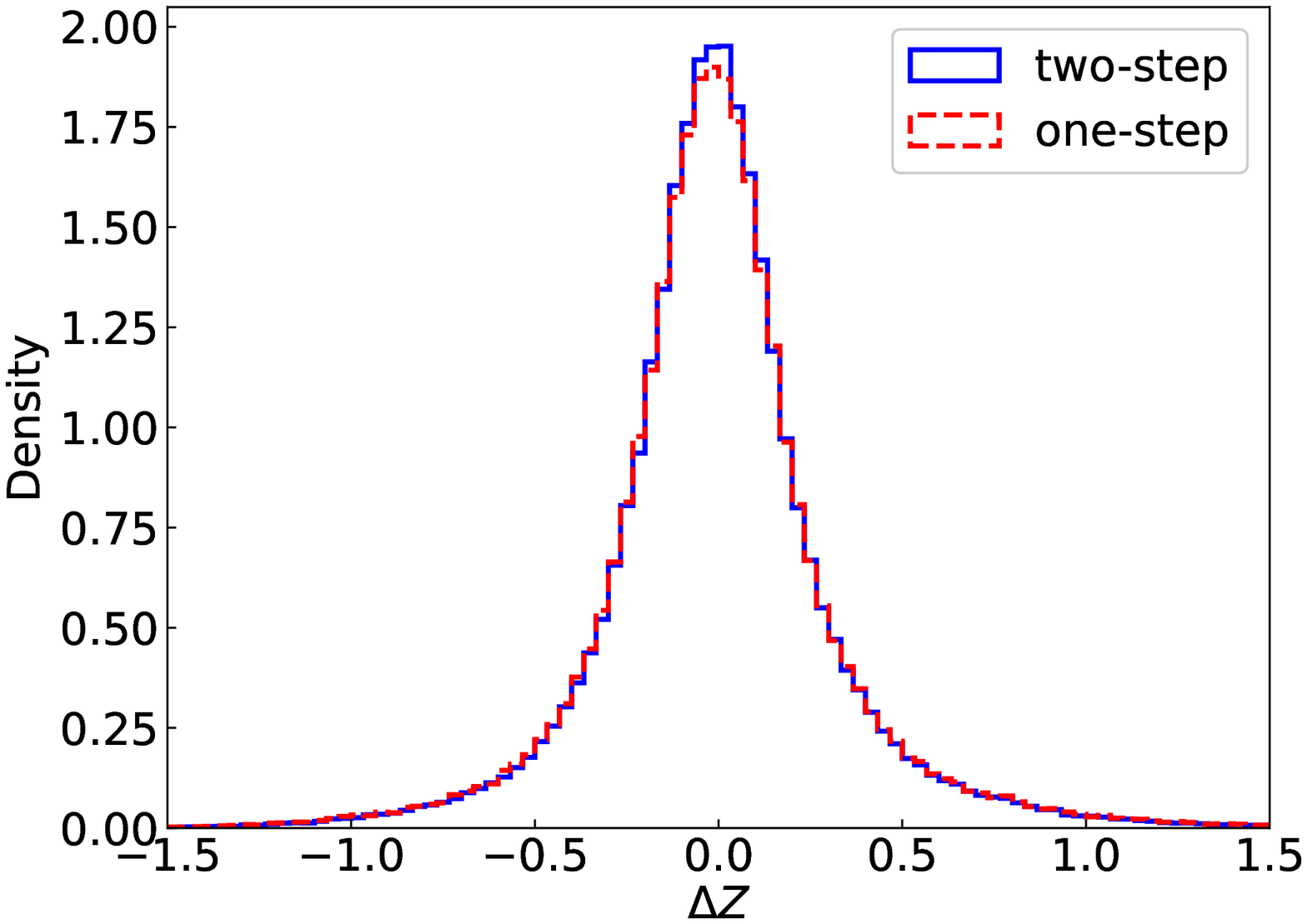}
                                  		\includegraphics[height=6cm,width=7cm]{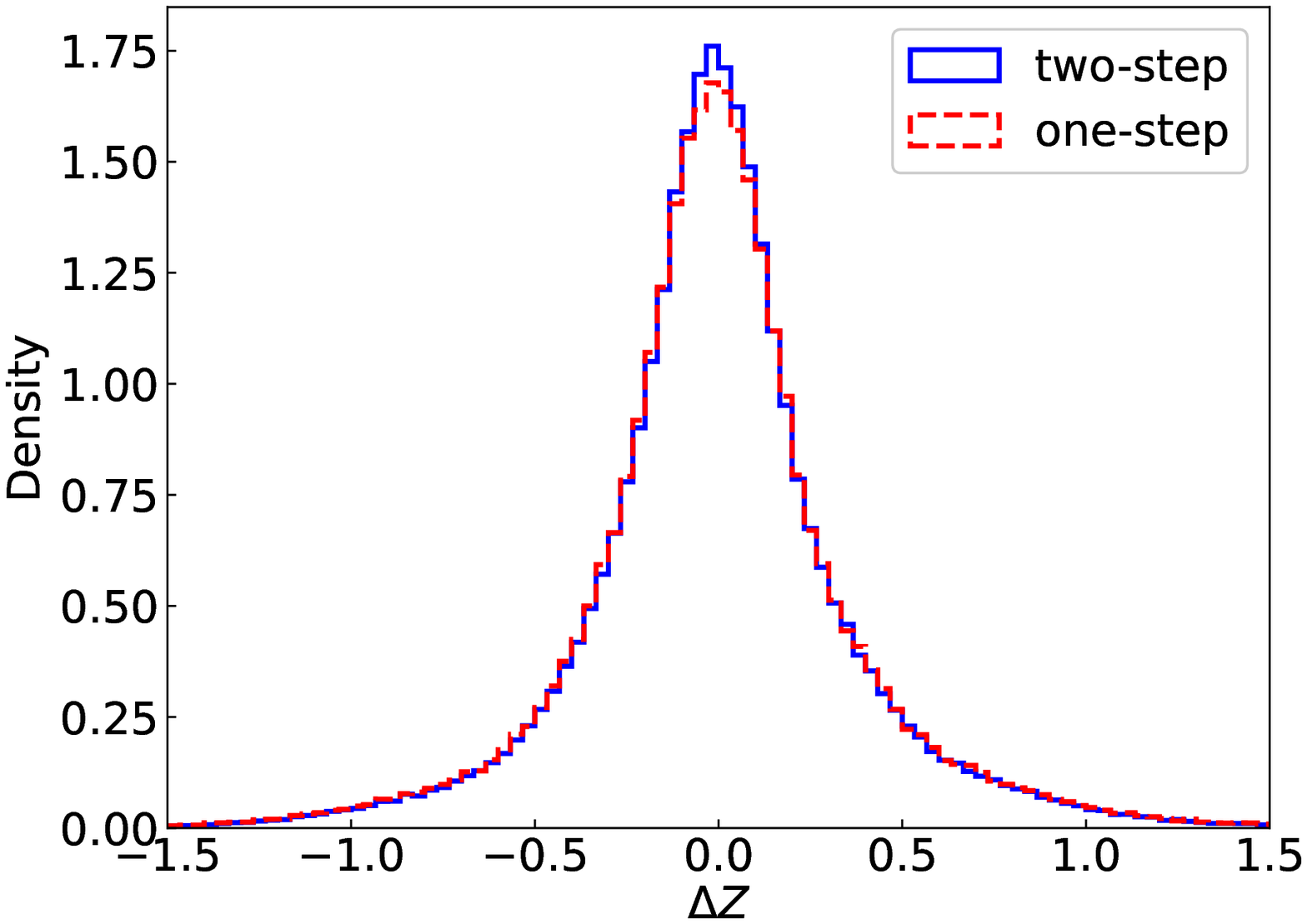}
                                  	}
                                  	\caption{The ${\Delta}{\mathrm{z}}$ distribution for the samples BSW (left panel) and BS\_W (right panel) with two-step models and one-step models during the range from -1.5 to 1.5, respectively.}
                                  	\label{fig6}
                                  \end{figure*}

\subsection{Application}
Based on the above experimental results for the known samples, we put forward a workflow of photometric redshift estimation as shown in Figure~7, which is applied to predict photometric redshifts of quasar candidates from BASS DR3. In Figure~7, the red rectangle boxes indicate data analysis and black parallelograms represent intermediate data or results. The workflow includes six regressors and one classifier. The detailed model information of this workflow is shown in Table~10.

\begin{figure*}
	\centering
	\includegraphics[height=22cm,width=18cm]{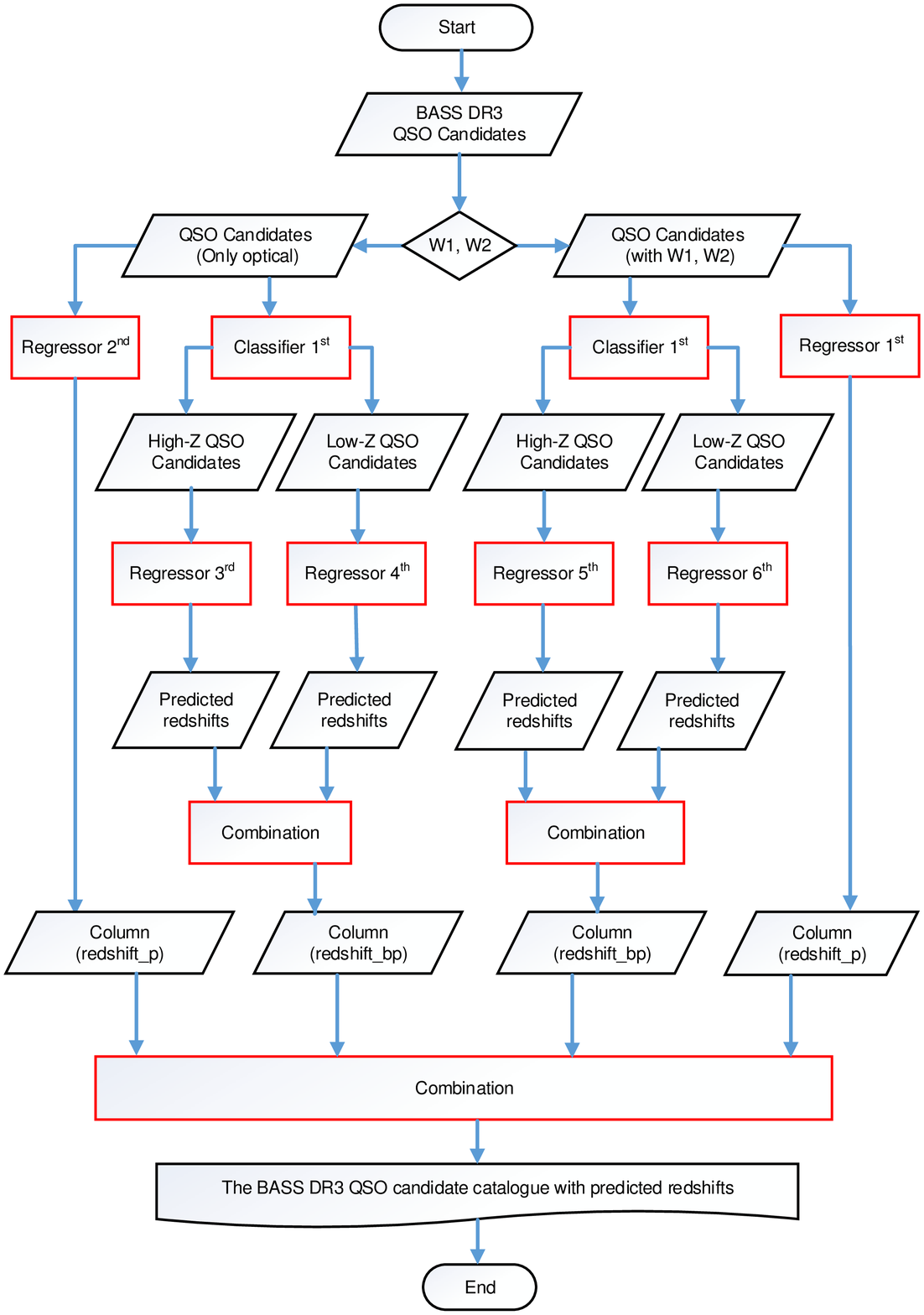}
	\caption{The photometric redshift estimation workflow.}
	\label{fig7}
\end{figure*}

             \begin{table*}
             	\begin{center}
             		\caption[]{The models used in photometric redshift estimation workflow.}
             		\begin{tabular}{rlllll}
             			\hline
             			Scheme     &Method &Input pattern  &  &Redshift range &$W1$, $W2$ \\
             			\hline
             			Regressor $1^{st}$     &CatBoost   &Pattern II &  &Full &Indispensable \\
             			Regressor $2^{nd}$     &CatBoost   &Pattern V &  &Full & Dispensable \\
             			Regressor $3^{rd}$     &CatBoost   &Pattern II &&Redshift $\ge$ 3.5 &Indispensable \\
             			Regressor $4^{th}$     &CatBoost   &Pattern II&&Redshift $<$ 3.5 &Indispensable\\
             			Regressor $5^{th}$     &CatBoost   &Pattern V & &Redshift $\ge$ 3.5 &Dispensable\\
             			Regressor $6^{th}$     &CatBoost   &Pattern V &&Redshift $<$ 3.5 &Dispensable\\
             			Classifier $1^{st}$    &CatBoost   &Pattern III & &(High redshift and low redshift) &Dispensable\\
             			\hline             		
             			\end{tabular}
             			\end{center}
             			\end{table*}

According to the work \cite{Li2021}, the BASS DR3 Sources were identified as Stars, Galaxies and Quasars by XGBoost. We consider all possible quasar candidates, the total number of which is 26 200 778. We adopt the workflow to estimate their photometric redshifts.

In Figure~7, BASS-DR3 quasar candidates are firstly divided into two samples. One sample contains only optical features, and the other sample contains optical and infrared features. For the candidates with optical and infrared features, we get predicted redshifts $redshift\_p$ by Regressor $^{1st}$, and these candidates are also classified into high redshift (high-Z) and low redshift (low-Z) subsamples. For the candidates classified as high-Z sources, we use Regressor $^{3rd}$ to estimate redshifts, while for candidates classified as low-Z sources, Regressor $^{4th}$ is used. Similarly, we obtain estimated redshifts of the candidates with only optical features. $redshift\_bp$ is predicted by two-step model while $redshift\_p$ is from one-step model. In the end, all predicted results are combined in a whole table. The link address is http://paperdata.china-vo.org/Li.Changhua/bass/bassdr3-quasar-z.hdf5. Table~11 lists 20 rows of predicted results, which is of great value for the further research on the characteristics and physics of these quasar candidates.

\begin{table*}
	\begin{center}
		\caption[]{The estimated redshifts of BASS DR3 quasar candidates, redshift$\_bp$ is predicted redshift by two-step model, redshift$\_p$ is predicted redshift by one-step model.}
		\begin{tabular}{lllcc}
			\hline
			id          &ra         & dec      &redshift$\_bp$  &redshift$\_p$ \\
			\hline
			95429001151 &133.48449872099638 &84.64760058245369 &3.612 &2.755 \\
			95375007162 &146.34468863159154 &84.19521072349899 &3.756 &2.967 \\
			95375009802 &146.70683096395808 &84.34011991887981 &3.853 &3.071 \\
			95376013790 &156.8204830082884 &84.55993969686726 &3.820 &3.462 \\
			95432000953 &156.68410784655092 &84.62164596197978 &3.770 &2.216 \\
			95433002683 &162.41293319964882 &84.75661154235415 &3.831 &3.252 \\
			95379009607 &172.10642294951734 &84.54070384254878 &4.290 &2.970 \\
			95435000350 &172.16026329686127 &84.59069887892409 &3.584 &2.644 \\
			95439001178 &204.45978967988097 &84.63147943935314 &3.703 &2.324 \\
			95440003112 &208.24299714603515 &84.74509303019737 &3.835 &2.349 \\
			95440003256 &207.3572024876584 &84.75161970723823 &3.657 &3.329 \\
			95441002154 &214.02787846450985 &84.7184000420548 &3.643 &2.777 \\
			95372005855 &128.4637617956736 &84.17660965013285 &3.690 &2.329 \\
			95372007632 &127.96938333672199 &84.25455113312614 &3.818 &3.120 \\
			95372008421 &129.9774226081903 &84.30134401378302 &3.753 &2.407 \\
			95373007697 &136.73203714550777 &84.23113544297041 &3.875 &2.745 \\
			95373009831 &138.71782967532423 &84.32491202628485 &3.725 &2.288 \\
			95374010575 &143.50234604548734 &84.37099267752663 &3.858 &3.136 \\
			95312008225 &151.1039377338386 &83.5936910539942 &3.716 &3.013 \\
			95312012685 &150.5329880180199 &83.78811573097289 &3.861 &1.869 \\
			95376009700 &153.0999380695831 &84.3246689896127 &3.738 &2.487 \\
			
				\hline
		\end{tabular}
	\end{center}
\end{table*}

In the work \cite{Li2021}, when applying both optical and infrared information with the same predicted results by binary and multiclass classifiers, the number of quasar candidates is 2 195 180, 1 500 099 ($P_{\rm Q}>0.75$), and 798 928 ($P_{\rm Q}>0.95$). In Figure~8, we show the number density distribution of quasar candidates as a function of photometric redshifts estimated by one- and two-step models. It is found from Figure~8 that there is some difference between one- and two-step models for the predicted redshifts of the quasar candidates. Table~12 lists the number of BASS DR3 quasar candidates with $P_{\rm Q}>0.95$ in different redshift ranges by the two models, which suggests that two-step model is more easy to find high redshift quasars than one-step model.

	\begin{figure*}
		\centering
		\includegraphics[bb=84 239 531 558,width=8cm,height=6.5cm]{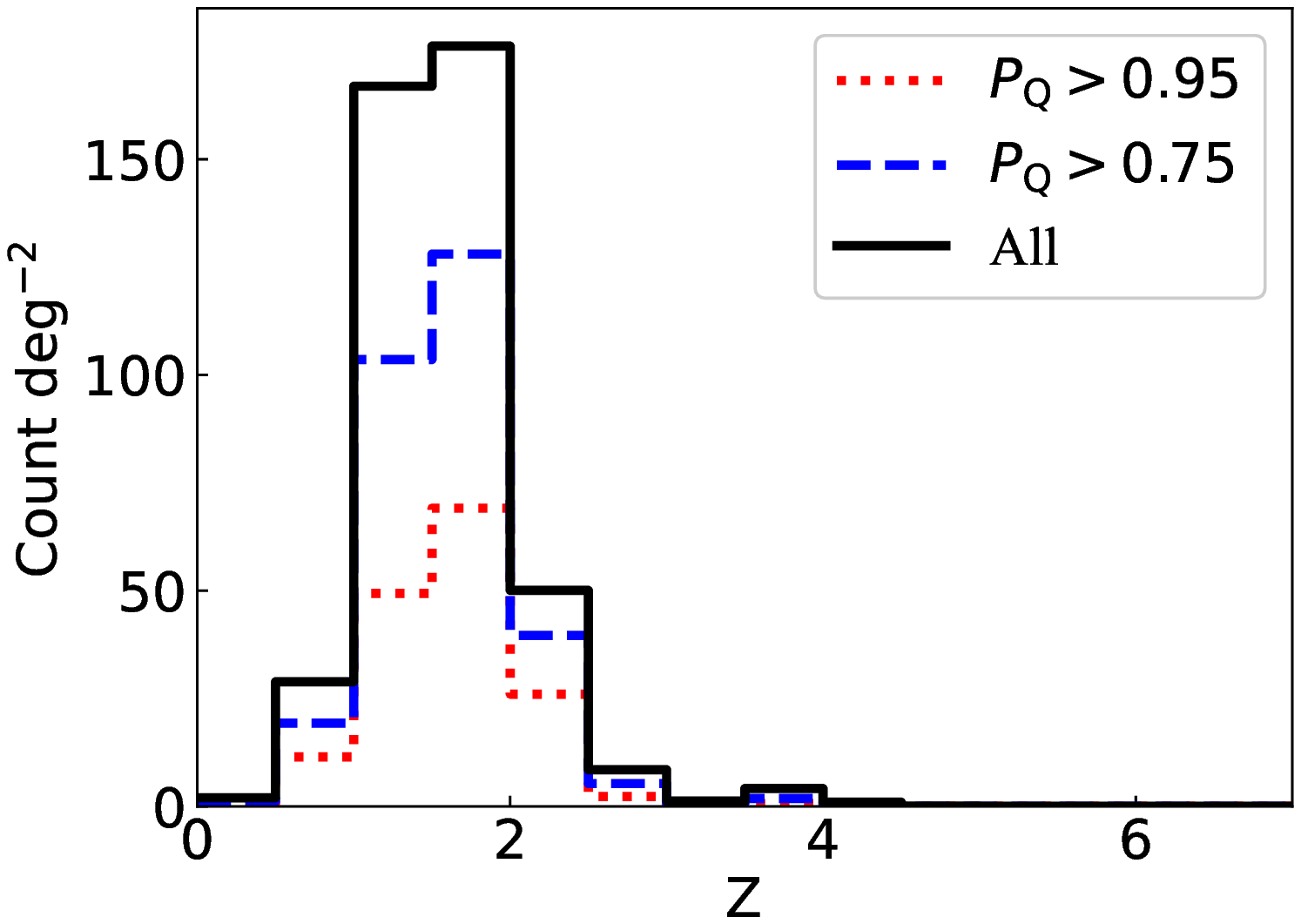}
		\includegraphics[bb=84 239 531 558,width=8cm,height=6.5cm]{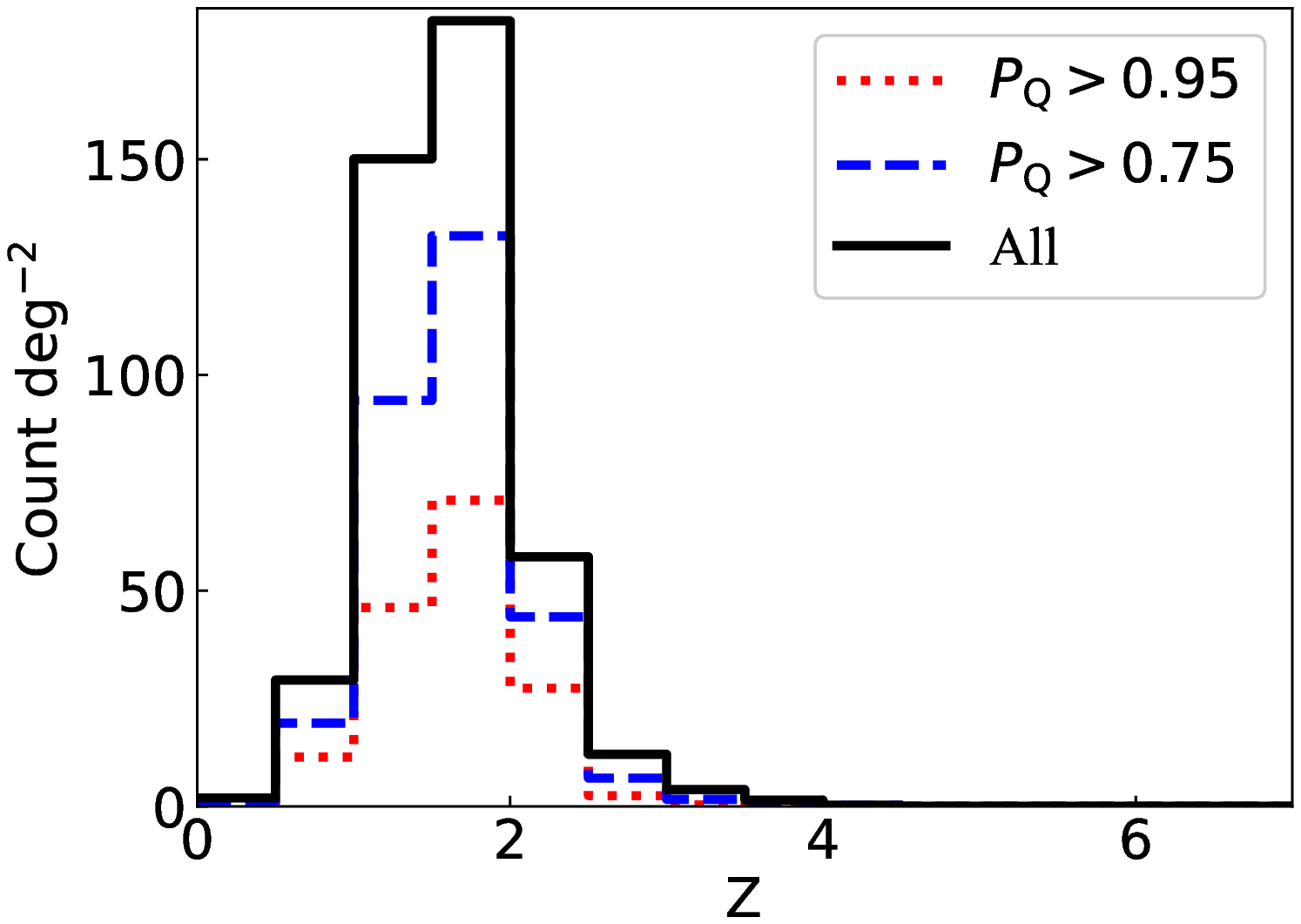}
		\caption{Left panel: the number density of quasar candidates as a function of photometric redshifts by two-step model; Right panel: the number density of quasar candidates as a function of photometric redshifts by one-step model (Regressor $1^{st}$). In both panels, the quasar candidates contain both optical and infrared information with the same predicted results by binary and multiclass classifiers for different probabilities ($P_Q>0.95 $: red dotted line, $P_Q>0.75$: blue dotted dash line, all: black line).}	
		\label{fig6}
	\end{figure*}

\begin{table*}
	\begin{center}
		\caption[]{The number of BASS DR3 quasar candidates with $P_{\rm Q}>0.95$ in different redshift ranges by two models.}
		\begin{tabular}{lllcc}
			\hline
			Model      & redshift$< 3.5$&$3.5\le$redshift$<4.5$     &$4.5\le$redshift$<5.5$ &redshift$\ge 5.5$ \\
			\hline
			 one-step&796078 &2822 &27 &1 \\
			 two-step&794990 &3817 &97 &24 \\
				\hline
		\end{tabular}
	\end{center}
\end{table*}

\section{Conclusions} \label{sec:conclusions}
 We present two schemes of machine-learning to predict photometric redshifts of quasars. We discuss the feature importance of different samples by XGBoost, CatBoost and Random forest. The optimal features and optimal model parameters of these three algorithms are chosen for different samples. Comparison of classification and regression of these three algorithms is performed. By contrast, CatBoost achieves the best performance among all classifiers and regressors for different samples considering effectiveness and efficiency. Moreover comparing the performance of two-step model with that of one-step model, two-step model is superior to one-step model. The two-step model is vital to select out high redshift quasar candidates. Therefore we put forward a photometric redshift workflow, in which CatBoost is taken as the core algorithm for classification and regression and two models are adopted. Then we utilize the workflow to predict photometric redshifts of all quasar candidates from BASS DR3. The predicted result will be of great help and reference for future research of quasars and can help LAMOST, DESI or other projects for follow up observation to find more high redshift quasars.

\section{Acknowledgements}
We are very grateful to the referee for his constructive suggestions and comments. This work is supported by National Natural Science Foundation of China (NSFC)(grant Nos. 11573019, 11803055, 11873066, 12133001, 11433005), the Joint Research Fund in Astronomy (U1531246, U1731125, U1731243, U1731109) under cooperative agreement between the NSFC and Chinese Academy of Sciences (CAS), the 13th Five-year Informatization Plan of Chinese Academy of Sciences (No. XXH13503-03-107) and the science research grants from the China Manned Space Project with NO. CMS-CSST-2021-A06. We would like to thank the National R\&D Infrastructure and Facility Development Program of China, ``Earth System Science Data Sharing Platform" and ``Fundamental Science Data Sharing Platform" (DKA2017-12-02-07). Data resources are supported by Chinese Astronomical Data Center (NADC) and Chinese Virtual Observatory (China-VO). This work is supported by Astronomical Big Data Joint Research Center, co-founded by National Astronomical Observatories, Chinese Academy of Sciences and Alibaba Cloud. This research has made use of BASS DR3 catalogue. BASS is a collaborative program between the National Astronomical Observatories of the Chinese Academy of Science and Steward Observatory of the University of Arizona. It is a key project of the Telescope Access Program (TAP), which has been funded by the National Astronomical Observatories of China, the Chinese Academy of Sciences (the Strategic Priority Research Program. The Emergence of Cosmological Structures grant no. XDB09000000), and the Special Fund for Astronomy from the Ministry of Finance. BASS is also supported by the External Cooperation Program of the Chinese Academy of Sciences (grant No. 114A11KYSB20160057). The BASS data release is based on the Chinese Virtual Observatory (China-VO). The Guoshoujing Telescope (the Large Sky Area Multi-object Fiber Spectroscopic Telescope, LAMOST) is a National Major Scientific Project built by the Chinese Academy of Sciences. Funding for the project has been provided by the National Development and Reform Commission. LAMOST is operated and managed by the National Astronomical Observatories, Chinese Academy of Sciences.

We acknowledgment SDSS databases. Funding for the Sloan Digital Sky Survey IV has been provided by the Alfred P. Sloan Foundation, the U.S. Department of Energy Office of Science, and the Participating Institutions. SDSS-IV acknowledges support and resources from the Center for High-Performance Computing at the University of Utah. The SDSS web site is www.sdss.org. SDSS-IV is managed by the Astrophysical Research Consortium for the Participating Institutions of the SDSS Collaboration including the Brazilian Participation Group, the Carnegie Institution for Science, Carnegie Mellon University, the Chilean Participation Group, the French Participation Group, Harvard-Smithsonian Center for Astrophysics, Instituto de Astrof\'isica de Canarias, The Johns Hopkins University, Kavli Institute for the Physics and Mathematics of the Universe (IPMU) /University of Tokyo, Lawrence Berkeley National Laboratory, Leibniz Institut f\"ur Astrophysik Potsdam (AIP), Max-Planck-Institut f\"ur Astronomie (MPIA Heidelberg), Max-Planck-Institut f\"ur Astrophysik (MPA Garching), Max-Planck-Institut f\"ur Extraterrestrische Physik (MPE), National Astronomical Observatories of China, New Mexico State University, New York University, University of Notre Dame, Observat\'ario Nacional / MCTI, The Ohio State University, Pennsylvania State University, Shanghai Astronomical Observatory, United Kingdom Participation Group, Universidad Nacional Aut\'onoma de M\'exico, University of Arizona, University of Colorado Boulder, University of Oxford, University of Portsmouth, University of Utah, University of Virginia, University of Washington, University of Wisconsin, Vanderbilt University, and Yale University.

\section{Data availability}
The predicted photometric redshifts for BASS-DR3 quasar candidates are saved in a repository and can be obtained by a unique identifier, part of which is indicated in Table~11. It is put in paperdata at http://paperdata.china-vo.org, and can be available with http://paperdata.china-vo.org/Li.Changhua/bass/bassdr3-quasar-z.hdf5.

\end{document}